\documentclass[a4paper,11pt]{article}
\usepackage[english]{babel}
\usepackage[utf8x]{inputenc}
\usepackage[T1]{fontenc}
\usepackage[a4paper,top=2cm,bottom=2cm,left=2cm,right=2cm,marginparwidth=2cm]{geometry}
\usepackage{multirow}
\usepackage{graphicx}
\usepackage{amsmath}
\usepackage{bbm,booktabs,microtype}
\usepackage{indentfirst}  
\usepackage{makecell}
\usepackage{threeparttable}
\usepackage{setspace}
\usepackage{adjustbox}
\usepackage{array}
\usepackage{subfig}
\usepackage{caption}
\usepackage{color, paralist}
\usepackage[table,xcdraw]{xcolor}
\usepackage{appendix}
\usepackage{natbib}
\usepackage{authblk}
\usepackage{amssymb,amsmath,bm,mathrsfs,makeidx,amsfonts,graphicx,amsthm,multirow}
\usepackage[figuresright]{rotating}
\usepackage{longtable}
\usepackage{algorithm,algorithmic} 
\newcolumntype{?}{!{\vrule width 1pt}}
\numberwithin{equation}{section}
\graphicspath{{plots/}}

\usepackage[pdftex,colorlinks=true,hypertexnames=false]{hyperref}
\definecolor{darkblue}{rgb}{0,0,.6}
\hypersetup{citecolor=darkblue,linkcolor=darkblue,urlcolor=darkblue}

\newcommand{\beqs}{\begin{eqnarray}}
\newcommand{\eeqs}{\end{eqnarray}}
\newcommand{\beqsn}{\begin{eqnarray*}}
\newcommand{\eeqsn}{\end{eqnarray*}}

\newtheorem{remark}{Remark}
\newtheorem{prop}{Proposition}
\newtheorem{lemma}{Lemma}
\usepackage{orcidlink}
\newtheorem{theorem}{Theorem}

\newtheorem{assumption}{Assumption}

\newcommand{\probP}{\text{I\kern-0.15em P}}

\begin{document}
\title{Uncertainty Learning for High-dimensional Mean-variance Portfolio}
\author[a]{Ruike Wu}
\author[b]{Yanrong Yang\orcidlink{0000-0002-3629-5803}}
\author[c]{Han Lin Shang\orcidlink{0000-0003-1769-6430}}
\author[a]{Huanjun Zhu\orcidlink{0000-0003-0575-4525}}
	
\affil[a]{Xiamen University}
\affil[b]{The Australian National University}
\affil[c]{Macquarie University}

\date{}

\maketitle
\begin{abstract}
Robust estimation for modern portfolio selection on a large set of assets becomes more important due to large deviation of empirical inference on big data. We propose a distributionally robust methodology for high-dimensional mean-variance portfolio problem, aiming to select an optimal conservative portfolio allocation by taking distribution uncertainty into account.     
With the help of factor structure, we extend the distributionally robust mean-variance problem investigated by \citet[][Management Science]{blanchet2022distributionally} to the high-dimensional scenario and transform it to a new penalized risk minimization problem. 
Furthermore, we propose a data-adaptive method to estimate the quantified uncertainty size, which is the radius around the empirical probability measured by the Wasserstein distance. 
Asymptotic consistency is derived for the estimation of the population parameters involved in selecting the uncertainty size and the selected portfolio return.
Our Monte-Carlo simulation results show that the chosen uncertainty size and target return from the proposed procedure are very close to the corresponding oracle version, and  the new portfolio strategy is of low risk. Finally, we conduct empirical studies based on S\&P index components to show the robust performance of our proposal in terms of risk controlling and return-risk balancing. 
\onehalfspacing
\smallskip
 
\end{abstract}

\onehalfspacing
 
\newpage
\section{Introduction}\label{sec:1}

The mean-variance portfolio selection problem proposed by \cite{markowitz1952portfolio} is one of the classical strategies in portfolio allocation. For a universe of $p$ stocks, the Markowitz mean-variance model tries to choose an allocation $w$ that minimizes portfolio risk while achieving the target expected return:
\begin{align}
\mathop{\min}_{w} \ w^\top \text{Var}_{\mathbb{P}^*}(r_t)w \label{model:classic}\\
\text{s.t} \; \ w^\top \textbf{1}_p = 1, \quad \text{E}_{\mathbb{P}^*}(w^\top r_t) \geq \overline{\alpha}, \notag
\end{align}
where $w$ is the portfolio weight, $r_t$ is $p$ dimensional assets return,  $\mathbb{P}^*$ is the probability measure underlying the distribution of $r_t$, $\text{E}_{\mathbb{P}^*}(\cdot)$ and $\text{Var}_{\mathbb{P}^*}(\cdot)$ are the expectation and variance under $\mathbb{P}^*$, $\overline{\alpha}$ is the target return, and $\textbf{1}_p$ is a $p$-dimensional column vector of ones. In practical application, however, we do not always have prior information about  population mean and covariance and hence rely instead on empirical versions of mean and covariance instead, which  usually deviate significantly  from the original \citep[see, e.g.,][]{michaud1989markowitz, jagannathan2003risk, ao2019approaching}. Additionally, model misspecification would result in the uncertainty in model and lead to inaccurate estimations.

These motivate the creation of the ``robust" version of the Markowitz model, which aims to consider the influence of disparities between $\mathbb{P}^*$ and its empirical counterpart. \cite{lobo2000worst} were the first to study the worst-case robust analysis with respect to  second-order moment uncertainty in the framework of the Markowitz model. \cite{goldfarb2003robust}  considered a vector/matrix distance for computing uncertainty set. \cite{pflug2007ambiguity} developed a Markowitz model with a Wasserstein distance-based ambiguity set and an additional value-at-risk type of constraint, but  in their study  the uncertainty size is exogenous. \cite{delage2010distributionally} examined  moment-based uncertainty regions in portfolio optimization and did not consider an optimal choice of the uncertainty set.  \cite{mohajerin2018data} introduced representations for the worst-case expectations in a Wasserstein-based uncertainty set and further applied their model to portfolio selection with various risk measures. A data-driven method for ambiguity size is also provided, but the choice deteriorates substantially with an increase in portfolio dimension.

Recently, \cite{blanchet2022distributionally}(hereafter, BCZ) proposed a distributionally robust mean-variance problem by using the Wasserstein distance-measured uncertainty set, which performs very well  in practical application. They showed the distributionally robust problem has a dual equivalent form that is a non-robust portfolio selection problem, and further provided data-driven methods for choosing  uncertainty size and the lowest target portfolio return involved in their model. However, in high-dimensional situations, due to the curse of dimensionality, the selected uncertainty size is often overestimated, leading to overly conservative portfolio allocation. Moreover, their theoretical results also fail under a high-dimensional framework. The selection of the uncertainty set is crucial for distributionally robust problems. 
\cite{zhang2023high} criticized the unnecessarily conservative allocation in  BCZ, and proposed a new way to choose  uncertainty size  using the subspace Wasserstein distance.  
However, their method requires sample splitting, resulting in a relatively small sample size for each sub-block used to estimate the population mean and covariance. 

In this paper, we consider a high dimensional distributionally robust mean-variance problem, which extends the model of BCZ to high dimensionality. To address  the curse of dimensionality, we impose the factor structure for observed asset returns, and the model uncertainty is quantified for common factors via the Wasserstein distance. We believe that the uncertainty of asset returns can largely be explained by common factors that drive the primary co-movements of asset returns, especially during significant macroeconomic changes, such as the global financial crisis and the outbreak of COVID-19.
Adding the factor structure for data in high-dimensional situations is quite popular and has become a frequently used workhorse (see, e.g., \cite{fan2013large}, \cite{ait17}, \cite{ding2021high}, \cite{fan2022}, etc.). We first demonstrate that the distributionally robust optimization with factor-based asset return can also be reformed to a non-robust risk-minimization problem with an additional regularization term. Similar to the BCZ model, their optimization in high dimensional situations also includes two key parameters, $\delta$ and $\rho$, which depict the ambiguity set size and the lowest target portfolio return given uncertainty set, respectively. Second, we provide  data-driven methods for determining the values of both $\delta$ and $\rho$. This is accomplished by adapting and extending the robust Wasserstein profile inference (RWPI) framework introduced by \cite{blanchet2019robust}. The core idea is to select  $\delta$ and $\rho$ such that the uncertainty set encompasses the true probability measure, and the feasible set of optimization includes the oracle mean-variance solution of \eqref{model:classic}, both at a specified confidence level. Third, since choosing procedures involve unknown population parameters, we provide  related estimation procedures, as well as corresponding theoretical consistency under some regularized conditions. Next, we conduct Monte Carlo simulations to evaluate the estimated "optimal" values of $\rho$ and $\delta$ obtained from the proposed selection procedures. These estimated values are quite close to the unknown oracle versions.  We also evaluate the performance of our high dimensional distrbituionally robust portfolio, which exhibits lower risk compared to that of BCZ. Finally, we study  empirical applications based on S\&P 500 index components, finding that over a long-term investment from 2000 to 2019, the proposed high dimensional distributionally robust portfolio achieves the lowest portfolio risk and highest Sharpe ratio (SR) among all considered strategies, including portfolios from BCZ.

Our paper also relates to the regularized portfolio allocation and high-dimensional portfolio allocation in finance. \cite{demiguel2009generalized} developed a general framework that solves the minimum variance problem (MVP) by adding the norm of the portfolio-weight vector such as $l_1$-norm and $l_2$-norm. \cite{fan2012vast} investigated MVP under gross-exposure constraint, where the $l_1$-norm of the weight vector is required to be bounded.  Based on MVP, \cite{fastrich2015constructing} studied the construction of an optimal sparse portfolio using a weighted Lasso penalty and a non-convex SCAD penalty. \cite{ho2015weighted} proposed regularizing the mean-variance problem with a weighted elastic net penalty, which can be motivated by a robust reformulation of the mean-variance criterion that directly accounts for parameter uncertainty. \cite{olivares2018robust} regularized the mean-variance problem with $p$-norm transaction costs and showed  equivalence between certain transaction costs and ellipsoidal robustification around the  mean. \cite{ding2021high} developed a unified minimum variance portfolio under statistical factor models in high-dimensional situations, with their approach also relying on properly integrating $l_1$ constraint on portfolio weights. \cite{brodie2009sparse} proposed adding $l_1$-norm penalty to the mean-variance problem and reforming it as a constrained least squares regression problem. \cite{ao2019approaching} translated the mean-variance optimization into a novel unconstrained regression representation in a high-dimensional situation and theoretically proved that the regularized portfolio based on $l_1$-norm regularization achieves the optimal mean-variance trade-off. However, their results rely on the Gaussian assumption and do not impose the budget constraint in \eqref{model:classic}. \cite{li2022synthetic} argued that the strategy of \cite{ao2019approaching} would not apply to real large-scale portfolio allocation since the response variable involves the inverse of the sample covariance matrix. Consequently, they recommended applying a covariance matrix estimator based on the Fama-French factor model and performing portfolio allocation via a synthetic regression.

This paper is organized as follows. Section~\ref{sec:2} introduces the basic high dimensional distributionally robust problem based on a factor model. The corresponding dual problem is derived. Section~\ref{sec:3} further studies the data-driven choice of required tuning parameters $\delta$ and $\rho$, the associated practical implementation of computing $\delta$ and $\rho$ is given in Subsection~\ref{sec:3.3}. Section~\ref{sec:5} presents an asymptotic theory. Section~\ref{sec:6} presents simulation results to show the superiority of our proposed method. Section~\ref{sec: empirical} develops some empirical applications based on S\&P 500 index components. Section~\ref{sec:8} concludes. The proof is included in the Appendix. 

In this paper, $\| A\|$, $\|A\|_1$, $\|A\|_F$, and $\| A\|_{\max}$ denotes the spectral norm, $L_1$ norm, Frobenius norm, and elementwise norm of a matrix A, defined respectively by $\|A\| = \lambda_{\max}^{1/2}(A^\top A)$, $\|A\|_1 = \max_{j}\sum_{i} |a_{ij}|$, $\|A\|_F = \text{tr}^{1/2}(A^\top A)$, and $\|A\|_{\max} = \max_{i,j}|a_{ij}|$, $\lambda_{\max}(A)$ ($\lambda_{\min}(A)$) denotes the maximum (minimum) eigenvalues of a matrix A and $tr(A)$ is the trace of a matrix A. If $A$ is a vector, $\|A\|$ is equal to the Euclidean norm. $\mathbb{P}^{*}, \mathbb{P}_{T}$, and $\mathbb{P}$ are true probability, empirical probability, and a general form of probability measures supported by $\mathbb{R}^{x}$ where $x$ takes different values in different situations.

\section{Methodology}\label{sec:2}

\subsection{MVP with Distributional Uncertainty}

We begin by reviewing the distributionally robust mean-variance model investigated by \cite{blanchet2022distributionally} (hereafter, BCZ). For $p$-dimensional  asset return $r_t$, the distributionally robust mean-variance portfolio is given by solving: 
\begin{align}
\mathop{\min}_{w} \mathop{\max}_{\mathbb{P} \in \mathcal{U}_\delta(\mathbb{P}_T)} \left\{ w^\top \text{Var}_{\mathbb{P}}(r_t)w \right\} \label{Blencht2022 model}
\\
\text{s.t.} \quad w^\top \textbf{1}_p = 1, \; \min_{\mathbb{P}\in \mathcal{U}_\delta(\mathbb{P}_T)} \text{E}_{\mathbb{P}}(w^\top r_t) \geq \rho \notag
\end{align}
where $w$ is portfolio allocation, $\text{E}_\mathbb{P}$ and $\text{Var}_{\mathbb{P}}$ take the expectation and variance under probability measure $\mathbb{P}$ respectively, $\mathbb{P}_T$ is the empirical probability measure, and $\mathcal{U}_\delta(\mathbb{P}_T) =\{\mathbb{P}:D_c(\mathbb{P},\mathbb{P_T})\leq \delta\}$ is the uncertainty set where $D_c(\cdot,\cdot)$ is the discrepancy between two probability measures based on a suitably defined Wasserstein distance
\begin{equation}
D_c(\mathbb{P},\mathbb{Q}) := \mathop{\inf} \left\{\text{E}_\pi\left[ c(U,V)\right] : \pi \in \mathcal{P}(\mathbb{R}^p\times \mathbb{R}^p), \; \pi_U = \mathbb{P},\; \pi_V = \mathbb{Q} \right\}, \label{eq: WPP define 1}
\end{equation}
where $\mathbb{P}$ and $\mathbb{Q}$ are two probability measures supported on $\mathbb{R}^p$, $c(U,V)$ is a lower semi-continuous cost function such that $c(U,U) = 0$ for any $U \in \mathbb{R}^p$, $\mathcal{P}(\mathbb{R}^p\times \mathbb{R}^p)$ is the set of joint probability distributions $\pi$ of $(U,V)$, and $\pi_U,\pi_V$ denote the marginal distribution of $U$ and $V$, respectively. Compared to the classic mean-variance problem \eqref{model:classic}, the distributionally robust
formulation \eqref{Blencht2022 model} introduces an artificial adversary $\mathbb{P}$ as a tool to account for the impact of the model uncertainty around the empirical distribution. Clearly, the solution of~\eqref{Blencht2022 model}  is robustly optimal, in the sense that it minimizes this upper bound on risk. The uncertainty model  involves two important parameters
$\delta$ and $\rho$, which control the size of the ambiguity set and the lowest acceptable target portfolio return, respectively. 

The min-max problem in~\eqref{Blencht2022 model} is complex and hard to solve. If the cost function takes the form of $c(U,V) = \|U - V\|_\iota^2$ where $\iota \geq 1$ is fixed, their Theorem~1 shows that the dual problem of~\eqref{Blencht2022 model} is a non-robust portfolio selection problem in terms of the empirical measure $\mathbb{P}_T$ with an additional regularization term:
\begin{align}
\mathop{\min}_{w} \sqrt{w^{\top} \text{Var}_{\mathbb{P}_T}(r_t)w} + \sqrt{\delta}||w||_q \label{blenchet2022 model2}
\\
\text{s.t.} \ w^\top \textbf{1}_p = 1, \quad\text{E}_{\mathbb{P}_T}(w^\top r_t) \geq \rho + \sqrt{\delta}||w||_q, \notag
\end{align}
where $1/q +1/\iota = 1$.
Both the objective function and corresponding feasible region are convex with respect to $w$ and, therefore, the new form~\eqref{blenchet2022 model2}  can be handled easily by convex optimization.

\subsection{Factor-structure MVP with Uncertainty}	

We notice that the distributionally robust problem \eqref{blenchet2022 model2} measures the model uncertainty centered around the empirical covariance matrix 
$\text{Var}_{\mathbb{P}_T}(r_t)$ and  empirical mean $\text{E}_{\mathbb{P}_T}(r_t)$. However, it is well known that the sample covariance matrix performs poorly when 
$p$ cannot be asymptotically negligible compared to the sample size $T$ due to the curse of dimensionality. Furthermore,  sample mean is not  a good estimation in the portfolio allocation problem, either, see a variety of paper investigating the minimum variance problem \citep{ledoit2017nonlinear, ding2021high, fan2022}. If the center of uncertainty set deviates significantly from the population counterpart, the corresponding allocation will tend to be unnecessarily conservative\footnote{In BCZ, the selected value of  $\delta$ ensures that the uncertainty set $\mathcal{U}_{\delta}$ includes the true probability measure at the given level, e.g. 95\%.}. Consequently, the model~\eqref{blenchet2022 model2} is not suitable for high-dimensional scenarios. In our empirical applications in Section~\ref{sec: empirical} and those of BCZ, the performance of the portfolio from~\eqref{blenchet2022 model2} is very close to the equal weighting strategy, which is a special case when $\delta = \infty$. Just as BCZ acknowledged, their approach
 fails when the number of stocks $p$ is not small compared with the sample size $T$, and the empirical probability measure is not even consistent when $p> T$. 

In this paper, we investigate the distributionally robust mean-variance strategy available in high-dimensional situations. To deal with the curse of dimensionality, we impose the factor structure on the asset returns, which is common in the financial field \citep{fama1993common,fama2015five,fan2013large}.   Following classic arbitrage pricing theory \citep{ross1976arbitrage} and  approximate factor model of \cite{Chamberlain83},  the asset return can be modeled as 
\begin{equation}
r_{it} = b_i^\top F_t + e_{it},\quad i = 1,\ldots,p; \quad t= 1,\ldots, T, \label{factor model}
\end{equation}
where $r_{it}$ is the excess return of asset $i$ at time $t$, 
$F_t$ represents the common factors that drive the co-movement of asset returns,  
$b_i$ is $K$ dimensional factor loading capturing the relationship between asset $i$ and common factors, $K$ is the number of common factors, 
and $e_{it}$ is the idiosyncratic error with zero mean. In matrix form, we can rewrite~\eqref{factor model} as 
\begin{equation*}
R = BF + E,
\end{equation*}
where $R$ is $p\times T$ excess return matrix, $r_{it}$ is the $i$\textsuperscript{th} row and $j$\textsuperscript{th} column element of $R$, $B = (b_1,\ldots,b_p)^\top$ is $p\times K$ factor loading matrix, $F = (F_1,\ldots,F_T)$, and $E = (e_1, \ldots, e_T)$ is $p\times T$ error matrix, and $e_t = \left( e_{1t},\ldots,e_{pt} \right)^{\top}$.

Based on factor structure, the population covariance matrix can be decomposed as $B\text{Var}_{\mathbb{P}^*}(F_t)B^\top + \text{Var}_{\mathbb{P}^*}(e_t)$. Thus, we consider the distributionally robust mean-variance problem available in high dimensional set-up as follows:
\begin{align}
\mathop{\min}_{w}  \left\{  \mathop{\max}_{\mathbb{P} \in \mathcal{U}_\delta(\mathbb{P}_T)}   w^\top \left[B\text{Var}_{\mathbb{P}}({F}_t)B^\top + \text{Var}_{\mathbb{P}^*}(e_t) \right] w  \right\} 
\label{our basic model}
\\
\text{s.t.} \quad w^\top \textbf{1}_p = 1, \; \min_{\mathbb{P}\in \mathcal{U}_\delta(\mathbb{P}_T)}\text{E}_{\mathbb{P}}(w^\top B {F}_t) \geq \rho. \notag
\end{align}
Unlike  model~\eqref{Blencht2022 model}, which imposes uncertainty on the entire asset return $r_t$, we impose the ambiguity set centered around the empirical measure on common factors. The number of factors $K$ is fixed and finite,  so it is reasonable for us to directly impose the uncertainty analysis on them.  For the high dimensional error term, we do not impose any uncertainty set for it in the interests of simplicity\footnote{It is possible to define a probability measure $\mathbb{P}_1$ such that $\text{Var}_{\mathbb{P}_1}(e_t)$ performs well under some regularized conditions.}.  
Furthermore,  based on factor structure, the target return constraint is simplified to  $\min_{\mathbb{P}} w^{\top} B\text{E}_{\mathbb{P}}(F_t) \geq \rho$ since the error term does not contribute to the expected excess return.
  

Similar to the model~\eqref{Blencht2022 model}, the problem~\eqref{our basic model} also involves two key parameters $\delta$ and $\rho$. Again, the parameter $\delta$ is generally interpreted as the power given to the adversary $\mathbb{P}$. The larger the value of~$\delta$, the more model ambiguity, and the less important is the information in the covariance matrix of $F_t$.\footnote{In the BCZ, the resulting optimal portfolio from~\eqref{Blencht2022 model} tends to be just equal allocations as $\delta\rightarrow\infty$.}  When $\delta = \infty$, the portfolio allocation from \eqref{our basic model} (assuming the target return constraint is slack) is the minimum variance portfolio\footnote{The solution from \eqref{model:classic} without target return constraint.} based on covariance $BB^\top + \text{Var}_{\mathbb{P}^*}(e_t)$. The parameter $\rho$ is generally interpreted as the lowest acceptable target return given the uncertainty set. In the classic mean-variance problem~\eqref{model:classic}, the target portfolio return is set to~$\overline{\alpha}$, and it is natural that the choice of $\rho$ is related to $\overline{\alpha}$. We note that even though the target portfolio return is often specified by the investors' preference and is very subjective, the choice of $\rho$ is still important as the optimizing regions of problem~\eqref{Blencht2022 model} and~\eqref{our basic model} may be empty if the value of $\rho$ is unsuitable\footnote{See more in Remark \ref{remark: no solution} and Simulation \ref{sec: simu port}.}.
Choosing $\rho = \overline{\alpha}$ is an aggressive plan and often leads to unsolved problems. It is more sensible to choose a smaller value than $\overline{\alpha}$ since we take the uncertainty set into consideration. 

In this paper, we choose the cost function $c(U,V) = \|U - V\|_\iota^2$ with $\iota \geq 1$. The following Theorem \ref{theorem 1} reformulates the min-max problem~\eqref{our basic model} to its dual form, which is much easier to solve in practice.

\begin{theorem}\label{theorem 1}
The primal formulation given in~\eqref{our basic model} is equivalent to the following dual problem:
\begin{align}
\mathop{\min}_{w} \left(\sqrt{w^{\top} B \text{Var}_{\mathbb{P}_T}(F_t) B^\top w}  + \sqrt{\delta} \| B^\top w\|_q \right)^2 + w^\top \Sigma_{e}^* w \label{our model reform}\\
\text{s.t.} \ w^\top \bm{1}_p = 1, w^{\top} B \text{E}_{\mathbb{P}_T}(F_{t}) \geq \rho + \sqrt{\delta}||B^\top w||_{q}, \notag
\end{align}
where $1/\iota + 1/q = 1$.
\end{theorem}

The transformation is conducted based on a duality argument shown in the Appendix. The objective function of~\eqref{our model reform} consists of the factor part (in the bracket) and the residual part. The additional regularization term $\sqrt{\delta}\| B^\top w\|_q$ is introduced by the uncertainty set and plays a penalizing role. The quantity $B^\top w$ can be regarded as the weight allocated to investing in the common factor in the framework of mean-variance allocation. 
Similar to regularization in portfolio allocation problems \citep[see, e.g.,][]{demiguel2009generalized, olivares2018robust}, the penalty $\sqrt{\delta}\| B^\top w\|_q$ provides robustness to the optimal portfolio from~\eqref{our model reform}. Additionally, although the cost function is chosen as $\|U-v\|_\iota^2$ here, the reformulation in Theorem~\ref{theorem 1} holds for any cost function of the form $c(U, V) = \|U-V\|^2$ where $\|\cdot\|$ is any given norm with a suitable dual, see more arguments in BCZ.

\begin{remark}\label{remark: no solution}
Note that both  constraints $\quad\text{E}_{\mathbb{P}_T}(w^\top r_t) \geq \rho + \sqrt{\delta}||w||_q$ in \eqref{blenchet2022 model2} and $w^\top B \text{E}_{\mathbb{P}_T}(F_t) \geq \rho + \sqrt{\delta}||B^\top w||_q$ in our \eqref{our model reform} may lead to the empty optimizing region given $\delta, \text{E}_{\mathbb{P}_T}(r_t), B$ and $\text{E}_{\mathbb{P}_T}(F_t)$. For example, in the case of $q = 2$ with $p = 1$, if $\sqrt{\delta} - \text{E}_{\mathbb{P}_T}(r_t) > 0, \rho > 0$, then there is no $w > 0$ satisfying   $\rho + \sqrt{\delta}|w| - \text{E}_{\mathbb{P}_T}(r_t) w \leq 0$. The larger the uncertainty set, the smaller the feasible region of 
$w$, since the target return constraint requires the portfolio return to exceed the threshold $\rho$ even when $E_{\mathbb{P}}(r_t)$ is very poor. Therefore, it is important to first check whether the optimizing region is empty or not before solving both dual form \eqref{blenchet2022 model2}  and  \eqref{our model reform}. Fortunately, the logic of choosing a value of $\rho$ in Subsection \ref{sec: choice of rho} automatically ensures a non-empty set with high probability, which simplifies the practical application of the method. See more discussion in Simulation \ref{sec: simu port}.
\end{remark}

\section{Uncertainty Quantification}\label{sec:3}

There are two key parameters $\delta$ and $\rho$ in our high dimensional distributionally robust model \eqref{our model reform}. In this section, we provide guidance on the choice of the size of the ambiguity set, $\delta$, as well as that of the worst mean return target, $\rho$. It is clear that the choice of both parameters should be informed by the data based on some statistical principles rather than being arbitrarily exogenous.
The logic of choosing values for $\delta$ and $\rho$ resembles that of BCZ and can be simply summarized as follows:  choose $\delta$ such that the uncertainty region is large enough to make the true probability measure become a plausible choice at a sufficiently high probability. After obtaining $\delta$, we further choose $\rho$ such that the feasible set of \eqref{our model reform} is large enough to at least make the oracle allocation of \eqref{model:classic} feasible at a given confidence level. 

\subsection{Choice of \texorpdfstring{$\delta$}{delta}}\label{sec: choice of delta}

The choice of uncertainty set $\delta$ is crucial for problem~\eqref{our basic model}. If $\delta$ is chosen as too large, there is too much model ambiguity, thus the information of factor covariance matrix is less relevant and the feasible region of problem \eqref{our basic model} is smaller. In this situation, the objective function degenerates to minimize $w^{\top} (BB^{\top} + \Sigma_{e}^{*})w$. The logic of choosing the value for $\delta$ is that it should be large enough such that the uncertainty set $\mathcal{U}_\delta(\mathbb{P}_T)$ contains the true probability measure~$\mathbb{P}^*$ with a given confidence level.  Discussing the probability measure and the related distance directly is an abstract task. However, if we can map the probability measure to some statistics derived from it, then we can study the less complex situation under which the statistic corresponding to the true probability measure is included in the set of statistics derived from all probability measures within the uncertainty set $\mathcal{U}_\delta(\mathbb{P}_T) \}$. Therefore, let us first introduce the following classic minimum variance problem based on common factors $F_t$  under probability measure $\mathbb{P}$:
\begin{align}
& \mathop{\min}_{\phi} \phi^\top  \text{Var}_{\mathbb{P}}({F}_t)\phi 
 \label{model: classic Factor MV problem}
\\ &
\text{s.t. \ } \phi^\top \textbf{1}_K = 1.
   \nonumber
\end{align}
where $\text{Var}_{\mathbb{P}}({F}_t)$ is the variance-covariance matrix of $F_t$ under probability measure $\mathbb{P}$. Note that $\text{Var}_{\mathbb{P}}({F}_t) = \left(\text{E}_{\mathbb{P}}({F}_t {F}_t^\top) - \text{E}_{\mathbb{P}}({F}_t)\text{E}_{\mathbb{P}}({F}_t^\top) \right)$.
Let  $\phi_{f,\mathbb{P}}$ solve above minimum variance problem \eqref{model: classic Factor MV problem}, $\phi_f^* := \phi_{f,\mathbb{P}^*}$ where ${\mathbb{P}}^*$ is the true probability measure, and $\Phi_{f,\mathbb{P}}$ be a set collecting all solutions $\phi_{f,\mathbb{P}}$\footnote{The solution for problem~\eqref{model: classic Factor MV problem} under probability measure $\mathbb{P}$ may be not unique.}. 

Unlike the procedure of choosing the uncertainty size based on the mean-variance problem  as described in \cite{blanchet2022distributionally}\footnote{In their study, the map of choosing  $\delta$ is given by the problem as follows:
\begin{align}
    \min \phi^{\top} \text{E}_{\mathbb{P}}(r_tr_t^\top) \phi,\ \text{s.t.}\ \phi^\top \text{1}_p = 1, \ \phi^\top \text{E}_{\mathbb{P}}(r_t) = \rho,
    \label{eq: blanchet model choose delta}
\end{align}
}, 
we choose the value based on the minimum variance problem with respect to common factors, because we only impose uncertainty on the factor part. 

\begin{remark}
In problem~\eqref{model: classic Factor MV problem}, we minimize the factor portfolio risk measured by $\text{Var}_{\mathbb{P}}(F_t)$ instead of $\text{E}_{\mathbb{P}}(F_tF_t^\top)$ to introduce the moment condition $\mu = \text{E}_{\mathbb{P}}(F_t)$ in function~\eqref{eq: RWP definition}, which finally contributes to the normal random variate in Theorem~\ref{Theorem: delta choose}. Additionally, the  approach to choosing based on the mean-variance problem \eqref{eq: blanchet model choose delta} studied by BCZ is less appealing in this paper, the results based on~\eqref{model: classic Factor MV problem} shown in Theorem~\ref{Theorem: delta choose} is much simpler than those in Theorem~2 of BCZ. 
\end{remark}

The following several technical assumptions are imposed to develop the  rule for choosing $\delta$.

\begin{assumption}
\label{assum: clt}
\begin{itemize}
\item[(a)]The common factor $\{F_t\}_{t=1}^T$ is a stationary and ergodic process satisfying $\text{E}_{\mathbb{P}^*}\left( ||F_t||_2^4 < \infty \right)$ for all $t$. 
\item[(b)] Let $\mu_f^{*}= \text{E}_{\mathbb{P}^*}(F_t)$, then $\max_{i\leq K} \mu_{f, i}^* < M$ for some positive constant M, where $\mu_{f,i}^*$ is the $i$\textsuperscript{th} element of $\mu_f^{*}$. Define  $$V_{g} := \lim_{T\rightarrow\infty}\frac{1}{T}\sum_{t=1}^T\sum_{t^\prime = 1}^TE_{\mathbb{P}^*}\left[\left( F_t -\mu_f^* \right)\left( F_{t^\prime} -\mu_f^* \right)^\top\right]$$ which is strictly positive definite, and the central limit theorem holds: $$\frac{1}{\sqrt{T}}\sum_{t=1}^T (F_t - \mu_f^*) \overset{d}{\rightarrow} Z_0:= N(\textbf{0}_K, V_g),$$
where $\textbf{0}_K$ is K-dimensional zero vector.
\end{itemize} 
\end{assumption}

\begin{assumption}\label{assum: probability}
For any deterministic matrix $\Lambda \in \mathbb{R}^{K\times K}$ and any deterministic vector $\daleth \in \mathbb{R}^K$ such that either $\Lambda \neq 0$ or $\daleth \neq 0$,
$$
\mathbb{P}^*\left(\| \Lambda F_t + \daleth \|_2>0 \right) >0.
$$
\end{assumption}

\begin{assumption}
Both $\Sigma_{f}^{*} = \text{Var}_{\mathbb{P}^*}(F_{t})$ and $\tilde{\Sigma}_{f}^{*} = \text{E}_{\mathbb{P}^*}(F_{t}F_{t}^{\top})$ are  positive definite, and the optimal weight $ \phi_f^*$ of problem \eqref{model: classic Factor MV problem} under $\mathbb{P}^*$ satisfies that $\|\phi_{f}^{*}\| \leq \overline{C}$ for some positive finite constant~$\overline{C}$.  
\label{assum:blenchet 3}
\end{assumption}

Assumption~\ref{assum: clt}(a) requires common factor $F_t$ to be a stationary and ergodic process with mild moment conditions. Assumption \ref{assum: clt}(b) assumes a central limit theorem for the partial sum of demeaned factors, which is common.
Assumption~\ref{assum: probability} holds by assuming that $F_t$ has a density, with a  similar condition found in \cite{blanchet2019robust} and BCZ. Assumption~\ref{assum:blenchet 3} indicates that  problem ~\eqref{model: classic Factor MV problem} has a unique solution  when $\mathbb{P} = \mathbb{P}^*$, thus  $\Phi_{f,\mathbb{P}^*} = \{\phi_f^*\}$, and therefore, there exist unique Lagrange multipliers $\lambda_{f}^*$  such that
\begin{equation}
2\left(\text{E}_{\mathbb{P}^{*}}(F_{t}F_{t}^{\top}) - \text{E}_{\mathbb{P}^{*}}(F_{t})\text{E}_{\mathbb{P}^*}(F_{t}^{\top})\right)\phi_{f}^{*} - \lambda_{f}^{*}\textbf{1}_{K} = 0. \label{eq: lagrange condition1}
\end{equation}

Now we further define that  $$\Gamma_\delta(\mathbb{P}_T) = \bigcup_{\mathbb{P}\in \mathcal{U}_\delta(\mathbb{P}_T)}\Phi_{f,\mathbb{P}}$$
collects all solutions of problem (\ref{model: classic Factor MV problem}) with $\mathbb{P} \in \mathcal{U}_\delta(\mathbb{P}_T)$. Thus, the true probability measure $\mathbb{P}^*$ being included in $\mathcal{U}_\delta(\mathbb{P}_T)$ is equivalent to  $\phi_f^*$ being included in $\Gamma_\delta(\mathbb{P}_T)$. As such, $\Gamma_\delta(\mathbb{P}_T)$ can be considered as a natural confidence region for $\phi_f^*$, and we choose $\delta_T^*$  for $\delta$  such that $\phi_f^*$ belongs to this region with a given confidence level, where
$$
\delta_{T}^{*} = \min\{\delta>0, \quad \mathbb{P}^*(\phi_f^*\in \Gamma_\delta(\mathbb{P}_T)) \geq 1-\alpha_0 \},
$$
where $1-\alpha_{0}$ is a given confidence level, for example, typically 95\%.

However, it is still difficult to compute $\delta_T^*$ directly by its definition. Therefore, we follow the idea of BCZ and apply the robust Wasserstein profile function (RWP) to provide a simpler representation. Since our uncertainty set is imposed on the common factor rather than on the return itself, and we use the minimum variance problem as a tool, the analysis differs  from that presented in BCZ. 
Before defining RWP, we first note that for any $\phi \in \Gamma_\delta(\mathbb{P}_{T})$, there exists Lagrange multiplier $\lambda_{f}$  such that
\begin{align}
2\left(\text{E}_{\mathbb{P}}(F_{t}F_{t}^{\top}) - \text{E}_{\mathbb{P}}({F}_t)\text{E}_{\mathbb{P}}({F}_{t}^{\top}) \right)\phi = \lambda_{f}\textbf{1}_{K}.\label{eq: FOC of classic problem}
\end{align}
By multiplying $\phi^{\top}$ on~\eqref{eq: FOC of classic problem}  and noting that $\phi^{\top} \textbf{1}_K = 1$, we obtain 
$$
\lambda_{f} = 2\phi^{\top} \left(\text{E}_{\mathbb{P}}(F_tF_t^\top) -\text{E}_{\mathbb{P}}({F}_t)\text{E}_{\mathbb{P}}({F}_t^\top) \right)\phi.
 $$
Then, the robust Wasserstein profile function can be defined as follows:
\begin{equation}
\overline{\mathcal{R}}_T(\phi,\lambda_{f}, \mu,\Sigma) := \inf\left\{D_c(\mathbb{P},\mathbb{P}_T): \; \lambda_{f}  = 2\phi^\top \left(\Sigma - \mu \mu^\top  \right)\phi, \; \mu = \text{E}_{\mathbb{P}}(F_t), \; \Sigma = \text{E}_{\mathbb{P}}(F_tF_t^\top)\right\} 
\label{eq: RWP definition}
\end{equation}
for $(\phi,\lambda_{f},\mu,\Sigma)\in \mathbb{R}^K \times \mathbb{R}\times \mathbb{R}^K \times \mathbb{S}_{+}^{K\times K}$, where $\mathbb{S}_{+}^{K\times K} $ is the set of all the symmetric positively semi-definite matrices, $D_c(\mathbb{P},\mathbb{P}_T)$ is the Wasserstein distance defined in (\ref{eq: WPP define 1}). Further, define
\begin{align}
\overline{\mathcal{R}}_T^*(\phi_f^*) := \inf_{ \lambda_{f} \in \mathbb{R}, \mu \in \mathbb{R}^K, \Sigma \in \mathbb{R}^{K\times K}}  \overline{\mathcal{R}}_T(\phi_f^*,\lambda_{f},\mu, \Sigma),
\end{align}
we can observe that the events 
\begin{equation}
\{ \mathbb{P}^* \in \mathcal{U}_\delta(\mathbb{P}_T)\} = \{ \phi_f^* \in \Gamma_\delta(\mathbb{P}_T)  \} \Rightarrow \{\overline{\mathcal{R}}_T^*(\phi_f^*) \leq \delta \}, \label{eq: event equiv 1}
\end{equation}
and for any given $\epsilon >0$,
\begin{equation}
\overline{\mathcal{R}}_T^*(\phi_f^*) \leq \delta + \epsilon \Rightarrow \phi_f^* \in \Gamma_\delta(\mathbb{P}_T).\label{eq: event equiv 2}
\end{equation}

Let us define 
$$
\tilde{\delta}_T^* = \inf \left\{\delta > 0,\quad \mathbb{P}^*( \overline{\mathcal{R}}_T^*(\phi_f^*) \leq \delta ) \geq 1 -\alpha_0 \right\}.
$$
Then, it follows from~\eqref{eq: event equiv 1} and~\eqref{eq: event equiv 2} that $\tilde{\delta}_T^* \leq \delta_T^* \leq \tilde{\delta}_T^* + \epsilon$ for arbitrary $\epsilon >0$, as such, we obtain 
$
{\delta}_T^* = \tilde{\delta}_T^*.
$
As a result, ${\delta}_{T}^{*}$ is given by the quantile corresponding to the $1-\alpha_{0}$ percentile of the distribution of $\overline{\mathcal{R}}_T^*(\phi_f^*)$.

In practice, it is difficult to handle $\overline{\mathcal{R}}_T^*(\phi_f^*)$ since it is a minimization problem in terms of the mean and variance in high-dimensional situations. As a replacement, we define an alternative statistic involving only the empirical variance-covariance, while producing the upper bound of $\delta_T^*$. Denote $\mathcal{S} = \text{E}_{\mathbb{P}_T}(F_tF_t^\top)$. To satisfy the condition in $\overline{\mathcal{R}}_T$, we set 
$
\lambda_T =  2\phi_f^{*\top}\left(\mathcal{S} - \mu_f^* \mu_f^{*\top} \right)\phi_f^* ,
$
where $\phi^*_f = \phi_{f,\mathbb{P}^*}$.  Let
$$
\mathcal{R}_T(\lambda_T, \mu_f^*, \mathcal{S}) := \overline{\mathcal{R}}_T(\phi_f^*,\lambda_{T}, \mu_f^*, \mathcal{S}).
$$
It is clear that $\mathcal{R}_T(\lambda_T, \mu_f^*,\mathcal{S}) \geq \overline{\mathcal{R}}_T^*(\phi_f^*)$, and thus  
$$
\mathcal{R}_T(\lambda_T, \mu_f^*,\mathcal{S}) \leq \delta \Rightarrow \overline{\mathcal{R}}_T^*(\phi_f^*) \leq \delta.
$$
Then, we have 
\begin{equation}\label{eq: delta bar > delta star}
\overline{\delta}_{T}^{*} := \inf \left\{\delta \geq 0: \mathbb{P}^*( {\mathcal{R}}_T(\lambda_T, \mu_f^*, \mathcal{S}) \leq \delta ) \geq 1 -\alpha_0 \right\} \geq \delta_T^*.
\end{equation}
Note that, by using the  choice of $\mathcal{S}$, $\mu_f^*$ and $\lambda_T$, $\mathcal{R}_T(\lambda_T, \mu_f^*,\mathcal{S})$ can be rewritten as
$$
\mathcal{R}_T(\lambda_T,\mu_f^*, \mathcal{S}) = \inf \left\{ D_c(\mathbb{P},\mathbb{P}_T): \text{E}_{\mathbb{P}}(F_t F_t^\top) = \mathcal{S},\; \text{E}_{\mathbb{P}}(F_t ) = \mu_f^{*}, \; 2\phi^{*\top}_f \left(\mathcal{S} -\mu_f^*\mu_f^{*\top}\right) \phi^*_f = \lambda_T \right\}.
$$
Based on the relationship in~\eqref{eq: delta bar > delta star}, we now use $\overline{\delta}_{T}^{*}$ as the choice of $\delta$. Following Theorem~\ref{Theorem: delta choose} provides the asymptotic properties of quantity $\mathcal{R}_T(\lambda_T,\mu_f^*, \mathcal{S})$.  If we can get access to the limit of $\mathcal{R}_T(\lambda_T,\mu_f^*,\mathcal{S})$, the $1-\alpha_0$ percentile of the limiting distribution can be used as an approximation of $\overline{\delta}_{T}^{*}$. 

\begin{theorem}\label{Theorem: delta choose}
Suppose Assumptions~\ref{assum: clt}-\ref{assum:blenchet 3} hold true. Then, we have
\begin{align}
T\mathcal{R}_T(\lambda_T,\mu_f^*,\mathcal{S}) \overset{d}{\rightarrow} L_0 := \mathop{\sup}_{\overline{\zeta}\in \mathbb{R}^K}\left\{ \overline{\zeta}^\top Z_0 - \mathop{\inf}_{\Gamma\in \mathbb{R}^{K \times K}} \left\{ \text{E}_{\mathbb{P}^*}\left( \left\|\Gamma F_t + \overline{\zeta}\right\|_q^2 \right)\right\} \right\}. 
\label{eq: Theorem 2 eq1}
\end{align}
where $Z_0 \sim N(0, V_g)$, $V_g$ is variance-covariance matrix of factors defined in Assumption~\ref{assum: clt}. Moreover, if $q = 2$, we have
\begin{align}
L_{0} = \frac{\| Z_0\|_2^2}{4\left( 1- \mu_f^{*\top}\tilde{\Sigma}_f^{*-1}\mu_f^*\right)}.\label{eq: Theorem 2 eq2}
\end{align}
\end{theorem}

Note that our random normal variate $Z_0$ is $K$-dimensional and its asymptotic variance is a simple long-run variance-covariance matrix of $F_t$. In contrast, the random normal variate in BCZ is $p$-dimensional and its asymptotic covariance is more complicated and includes the unknown optimal weight\footnote{The asymptotic variance-covariance matrix of corresponding random variate in BCZ is $\lim_{T\rightarrow\infty} \text{Var}_{\mathbb{P}^*}(T^{-1/2}\sum_{t=1}^Tg(r_t))$ where $g(x) = x + 2\left(xx^\top \phi_{mv}^* - \phi_{mv}^{*\top} xx^\top \phi^{*}_{mv}\textbf{1}_p\right)/\lambda_1^*$, $\phi_{mv}^*$ and $\lambda_1^*$ are optimal weight of mean-variance problem based on $r_t$ and corresponding Lagrange multiplier.}. Furthermore,  the denominator in~\eqref{eq: Theorem 2 eq2} is defined based on the low-rank common factor instead of asset return.

When $q \neq 2$, $L_0$ does not have an explicit expression, and in this situation, using the inequalities that $\| x\|_q^2 \geq \|x\|_2^2$ if $q <2$ and $K^{1/2-1/q }\| x\|_q^2 \geq \|x\|_2^2$ if $q > 2$, we can find a stochastic upper bound of $L_0$ that can be explicitly expressed. We note that in the BCZ empirical analysis, they discussed that $q = 1$ leads to some volatile results, see their discussions 5.3. As such, in what follows, we focus on the case of $q = 2$ where $L_0$ has an explicit expression. This is the simplest situation and may be more preferred by practitioners.

\subsection{Choice of \texorpdfstring{${\rho}$}{rho}}
\label{sec: choice of rho}

Once $\delta$ has been chosen, the next step is to choose the value of $\rho$. Given $\delta$, if $\rho$ is chosen too large, the feasible region of problem~\eqref{our basic model} becomes smaller and may even become empty. Consequently, ideally, $\rho$ should chosen so that the feasible domain has at least one point as much as possible. Specifically, we select the value for $\rho$ such that we do not rule out the inclusion $w_{mv}^{*}\in \mathcal{F}_{\delta,\rho}(T)$ with a given confidence level $1-\epsilon$, where $w^*_{mv}$ is the optimal solution of problem~\eqref{model:classic} under $\mathbb{P}^{*}$ and
$$
\mathcal{F}_{\delta,\rho}(T) = \left\{ w\in \mathbb{R}^p: w^\top\textbf{1}_p = 1, \; \min_{\mathbb{P}\in\mathcal{U}_{\delta}(\mathbb{P}_T)}\text{E}_{\mathbb{P}}(w^\top B F_t) \geq \rho
\right\}
$$
is the feasible set of problem~\eqref{our basic model}.

Let $\rho = \overline{\alpha} - \sqrt{\delta}\|B^\top w_{mv}^*\|_qv_0$, then choosing the value for $\rho $ is equivalent to choosing the value for~$v_0$.  By applying the proposition A.1.\footnote{Proposition A.1 of BCZ says that, for $c(u,v) = \|u-v\|_\iota^2$, we have $$\min_{\mathbb{P}\in\mathcal{U}_{\delta}(\mathbb{P}_T)}\text{E}_{\mathbb{P}}(w^\top r_t) = \text{E}_{\mathbb{P}_T}(w^\top r_t) - \sqrt{\delta}\|w \|_q,$$ 
where $1/\iota + 1/q = 1.$} of BCZ, it follows   that $w_{mv}^* \in \mathcal{F}_{\delta,\rho}(T)$ if and only if 
$$
w_{mv}^{*\top}B \text{E}_{\mathbb{P}_T}(F_t) - \sqrt{\delta}||B^\top w_{mv}^*||_q \geq \overline{\alpha} - \sqrt{\delta}||B^\top w^*||_qv_0.
$$
It is evident that if $w^*_{mv}$ is included in the feasible region $\mathcal{F}_{\delta,\rho}(T)$, it should satisfy the conditions required in $\mathcal{F}_{\delta,\rho}(T)$. Notice that $\overline{\alpha} = w_{mv}^{*\top} \text{E}_{\mathbb{P}^*}(r_t) = w_{mv}^{*\top} B\mu_f^*$, $\mu_f^{*} = \text{E}_{\mathbb{P}^*}(F_t)$, thus we can rewrite the previous inequality as 
\begin{align}
{w_{mv}^{*\top}B\left(\text{E}_{\mathbb{P}_T}(F_t) - \text{E}_{\mathbb{P}^*}(F_t)\right)} \geq  ||B^\top w_{mv}^*||_q \sqrt{\delta}(1-v_0). \label{eq: alpha select ineq}
\end{align}
It is evident that we can choose $1 - v_{0}$ sufficiently negative such that the inequality (\ref{eq: alpha select ineq}) holds at the given confidence level $1-\epsilon$. Once $v_0$ has been determined, we finish choosing $\rho$ via the mapping $\rho  = \overline{\alpha} - \sqrt{\delta}||B^\top w_{mv}^*||_q v_0$. 

Based on Assumption~\ref{assum: clt}, it follows that
\begin{align}
(w^{*\top}_{mv}B) \frac{1}{\sqrt{T}}\sum_{t=1}^T\left(F_t - \mu_f^* \right)
\overset{d}{\rightarrow} Z_1 \sim N(0, w^{*\top}_{mv}B V_g B^\top w^*_{mv}), \label{eq: clt for rho}
\end{align}
where $V_g$ is defined in Assumption~\ref{assum: clt}.
As a result, the value of $1-v_{0}$ is given by $T^{-1/2}\mathcal{A}/\sqrt{\delta}\|B^\top w_{mv}^*\|_q $ where
$\mathcal{A}$ is the $\epsilon$ quantile of $Z_1$. The final value of $\rho$ is given by the maximum value of $\overline{\alpha} - \sqrt{\delta}||B^\top w_{mv}^*||_q v_0$. Again, we focus on the situation where $q = 2$ for simplicity.

\subsection{Implementation}\label{sec:3.3}

In  a practical implementation we generally do not know the population quantities in advance. As replacements, in order to obtain the optimal choice of $\delta$ and $\rho$, consistent estimators for population parameters are required. From Sections~\ref{sec: choice of delta} and~\ref{sec: choice of rho}, the estimations of $V_g$, $\mu_f^{*\top}\tilde{\Sigma}_f^{*-1}\mu_f^*$, $w_{mv}^{*\top} BV_g B^\top w_{mv}^*$ and $\|B^\top w_{mv}^*\|_2$ are required while calculating the values of $\delta$ and $\rho$.

The factor loading and common factors are estimated by solving 
\begin{equation}
\min_{B \in \mathbb{R}^{p \times K}, F_t\in \mathbb{R}^K} \sum_{t=1}^T \| r_t - B F_t \|_2^2 \label{eq: PCA Obj}
\end{equation}
subject to the normalization for identification
\begin{equation}
\frac{1}{T} F F^{\top} = \bm{I}_{K}, \ \text{and\; } \frac{1}{p} B^\top B \text{\ is diagonal.} \label{eq: PCA constraint}
\end{equation}
It was shown that $\widehat{F} = \sqrt{T}eig_{K}\left( R^\top R \right)$ and $\widehat{B} = T^{-1}R\widehat{F}$ where $eig_K(M)$ takes the eigenvectors of $M$ corresponding to its K largest eigenvalues (see, e.g., \cite{bai2002determining}, \cite{fan2013large} and \cite{fan2018aos}). Then,  the term $\mu_f^{*\top} \tilde{\Sigma}_f^{*-1}\mu_f^*$ can be easily estimated by its sample version 
$\widehat{\mu}_f^\top \widehat{\tilde{\Sigma}}_f^{-1}\widehat{\mu}_f $, where $\widehat{\mu}_f = T^{-1}\sum_{t=1}^T \widehat{F}_t$ and $\widehat{\tilde{\Sigma}}_f = T^{-1}\sum_{t=1}^T \widehat{F}_t\widehat{F}_t^\top $. 

Once we obtain $\widehat{B}$ and $\widehat{F}$, we can estimate error terms by $\widehat{e}_t = r_t - \widehat{B}\widehat{F}_t$, and thus the adaptive sparse covariance matrix estimation of error term $\widehat{\Sigma}_{e,sp}$ is given by 
\begin{align}
\widehat{\Sigma}_{e,sp,ij} = \left\{\begin{array}{cc}
   s_{ii}  &  i = j \\
   \rho_{ij}(s_{ij})  & i \neq j
\end{array} \right. ,
\label{eq: estimated sparse cov}
\end{align}
where $s_{ij} = \frac{1}{T}\sum_{t=1}^{T} \widehat{e}_{it}\widehat{e}_{jt}$, $\rho_{ij}(\cdot)$ is shrinkage function that satisfies that  for all $z \in R$, $\rho_{ij}(z) = 0$ when $|z| \leq \tau_{ij}$ and $|\rho_{ij}(z)-z|\leq \tau_{ij}$, $\tau_{ij}$ is the threshold, which is set to $C\sqrt{\widehat{\theta}_{ij}}\omega_T$ where $C$ is some positive constant, $\widehat{\theta}_{ij} = \frac{1}{T}\sum_{t=1}^T\left( \widehat{e}_{it}\widehat{e}_{jt} - {s}_{ij} \right)^2$ and $\omega_{T} = \sqrt{1/p} + \sqrt{\log p/T}$.
As a result, the estimated population covariance matrix of excess return based on factor structure is given by:
\begin{equation}
\widehat{\Sigma}_{r} = \widehat{B}(\widehat{\tilde{\Sigma}}_f -\widehat{\mu}_f\widehat{\mu}_f^\top)\widehat{B}^\top + \widehat{\Sigma}_{e,sp}.\label{eq: poet estimator}
\end{equation}
Furthermore, with $\widehat{\Sigma}_r, \widehat{B}$, and $\widehat{\mu}_f$   in hand,  we can obtain the estimator $\|\widehat{B}^\top \widehat{w}_{mv}\|_2$ where $\widehat{w}_{mv}$ is the optimal solution of problem \eqref{model:classic} using $\widehat{\Sigma}_r$ and $\widehat{B}\widehat{\mu}_f$ instead of population counterparts.

We turn to the estimation of the long-run variance-covariance matrix $V_g$ of $F_t$, which plays a vital role in the procedure of selecting both $\delta$ and $\rho$.  Recall that $$V_g = \lim_{T\rightarrow\infty}\frac{1}{T}\sum_{t=1}^T\sum_{t^\prime = 1}^{T}\text{E}_{\mathbb{P}^*}\left[\left( F_t -\mu_f^* \right)\left( F_{t^\prime} -\mu_f^* \right)^\top\right],$$ by changing of variables, can be alternatively expressed as
$
V_g = \lim_{T\rightarrow\infty} \overline{V}_{g,T},
$ with $\overline{V}_{g,T} =  \sum_{j = -T +1}^{T-1}\mathcal{C}_T(j)$, where 
$$
\mathcal{C}_T(j) = \frac{1}{T}\sum_{t=j+1}^T \text{E}_{\mathbb{P}^*}\left[\left( F_t -\mu_f^* \right)\left( F_{t-j} -\mu_f^* \right)^\top\right]    \text{for}\ j\geq 0,
$$
and $\mathcal{C}_T(j) = \mathcal{C}_T(-j)$ for $j< 0 $ since $F_t$  stationary process.
The standard method to come up with a consistent estimator $\widehat{\Psi}_T$ is to apply the famous heteroskedasticity and autocorrelation (HAC) kernel estimation, which is given by 
\begin{align}
\widehat{V}_{g,T} = \widehat{\mathcal{C}}_T(0) + 2\sum_{j=1}^{T-1}k\left(\frac{j}{q_T}\right)\widehat{\mathcal{C}}_T(j), 
    \label{eq: HAC estimator}
\end{align}
where $k(\cdot)$ is real-valued kernel function satisfying Assumption \ref{assum: Vg} below,  $q_T$ is the bandwidth, and
\begin{align*}
\widehat{\mathcal{C}}_T(j) = \frac{1}{T}\sum_{t=j+1}^T\left(\widehat{F}_t -\widehat{\mu}_f \right)\left(\widehat{F}_{t-j} -\widehat{\mu}_f \right)^\top     \text{for} \ j\geq 0.
\end{align*}
Note that the HAC estimation $\widehat{V}_{g,T}$ is defined on estimated factors~$\widehat{F}_t$ since $F_t$ is unobservable in our set-up. Once we have estimated $V_g$, we can immediately estimate $w_{mv}^{*\top} BV_g B^\top w_{mv}^*$ by $\widehat{w}_{mv}^{\top} \widehat{B}\widehat{V}_g \widehat{B}^\top \widehat{w}_{mv}$.

Finally, we present an algorithm documenting the detailed procedure for choosing values for~$\delta$ and~$\rho$.

\begin{algorithm}[H]
\caption{\label{algonew delta}{Choice of $\delta$ and $\rho$}}	
\begin{algorithmic}
\STATE 1.  Calculate the estimation of  factor loadings and common factors by  $\widehat{F} = \sqrt{T}eig_K(R^\top R)$ and $\widehat{B} = T^{-1}R\widehat{F}$, where $eig_K(M)$ takes the eigenvectors of M corresponding to its $K$ largest eigenvalues.

\STATE 2. Compute $\widehat{\mu}_f = T^{-1}\sum_{t=1}^T \widehat{F}_t$,    $\widehat{\Sigma}_r = \widehat{B}(I_K - \widehat{\mu}_f\widehat{\mu}_f^\top )\widehat{B}^\top + \widehat{\Sigma}_{e,sp}$ and $\widehat{V}_{g,T}$ by \eqref{eq: HAC estimator} 
where $\widehat{\Sigma}_{e,sp}$ is sparse error covariance estimator defined in \eqref{eq: estimated sparse cov}.
  
\STATE 3. The choice of $\delta$ is given by $1/T$ times the $1-\alpha_0$ (i.e. 95\%) quantile of $\widehat{L}_0$ where
$
\widehat{L}_0 =  {\| \widehat{Z}_0\|_2^2}/{4\left( 1- \widehat{\mu}_f^\top \widehat{\mu}_f\right)}
$
and $\widehat{Z}_0 \sim N(0,\widehat{V}_{g,T})$.
  
\STATE 4.  Compute $\widehat{w}_{mv}$ which is the estimation  of $w^*_{mv}$ by replacing population covariance matrix and mean with $\widehat{\Sigma}_r$ and $
\widehat{B}\widehat{\mu}_f$ in \eqref{model:classic}.

\STATE 5. Compute $\epsilon$ (i.e., 5\%) quantile of normal variate with zero mean and variance $\widehat{w}_{mv}^\top \widehat{B}\widehat{V}_{g,T} \widehat{B}^\top \widehat{w}_{mv}$, denote it as $\widehat{\mathcal{A}}$.

\STATE 6. The choice of $\rho$ is given by $\overline{\alpha} - \left(\sqrt{\delta_c} \|\widehat{B}^\top \widehat{w}_{mv} \|_2 - T^{-1/2}\widehat{\mathcal{A}} \right)$, where $\overline{\alpha}$ is the target portfolio return in classic mean-variance problem \eqref{model:classic} and $\delta_c$ is the selected value of $\delta$ in step 3.
\end{algorithmic}
\end{algorithm}

\section{Theoretical Guarantee of Parameter Estimates}\label{sec:5}
We note that again, the calculations of $\delta$ and $\rho$ from Subsections \ref{sec: choice of delta} and \ref{sec: choice of rho} require the knowledge of  $V_g$, $\mu_f^{*\top}\tilde{\Sigma}_f^{*-1}\mu_f^*$, $w_{mv}^{*\top} BV_g B^\top w_{mv}^*$, and $\|B^\top w_{mv}^*\|_2$. In this section, we further provide the theoretical guarantee for the corresponding estimations proposed in Subsection \ref{sec:3.3}.

\subsection{Basic Assumptions}
The following basic technique assumptions ensure that our estimators, developed in the previous section, are asymptotically consistent with corresponding true values.

\begin{assumption}
All the eigenvalues of the $K \times K$ matrix $p^{-1}B^\top B$ are bounded away from both~0 and~$\infty$ as $p \rightarrow \infty$.\label{assum: loading}
\end{assumption}

\begin{assumption}
\begin{itemize}
\item[(a)] $\{e_t,F_t\}$ is strictly stationary. In addition, $E_{\mathbb{P}^*}(e_{it}) = E_{\mathbb{P}^*}(e_{it}F_{jt}) = 0$ for all $i\leq p, j\leq K$ and $t\leq T$.
\item[(b)] There are constants $c_1, c_2 > 0$ such that $\lambda_{\min}(\Sigma_e^*) > c_1$, $\left\vert \left\vert \Sigma_e^* \right\vert\right\vert_1 < c_2$ and $\mathop{\min}_{i\leq p,j\leq p} \text{Var}_{\mathbb{P}^*}(e_{it}e_{jt}) > c_1$.
\item[(c)] 
Let $m_\varsigma=\max_{i\leq p}\sum_{j\leq p}|\Sigma_{e,ij}^*|^{\varsigma}$
for some $\varsigma \in [0,1)$ satisfying that $\omega_T^{1-\varsigma} m_\varsigma = o(1)$ where $\omega_T = \sqrt{1/p} + \sqrt{\log p/T}$.
  
\item[(d)] There are $y_1, y_2 > 0$ and $\beta_1,\beta_2 > 0$ such that for any $s>0$, $i\leq p$ and $j \leq K$, $P^*(\left\vert e_{it} > s\right\vert) \leq \exp\{-(s/\beta_1)^{y_1} \}$ and  $P^*\left(\left\vert F_{jt} > s\right\vert\right) \leq \exp\{-(s/\beta_2)^{y_2} \}$.
\end{itemize}
\label{assum: error term}
\end{assumption}

\begin{assumption}
There exists $y_3 > 0$ such that $3y_1^{-1} + 1.5y_2^{-1} + y_3^{-1} > 1$, and $C > 0$ satisfying, for all $ t \in \mathbb{Z}^+$, $\alpha(t) \leq \exp(-Ct^{y_3})$, where
$\alpha$ is mixing coefficient defined by $\sup_{A\in \mathcal{F}_{-\infty}^0,B\in \mathcal{F}_{T}^\infty}\left| P^*(A)P^*(B) - P^*(AB) \right|$, $\mathcal{F}_{-\infty}^0$ and $\mathcal{F}_{T}^\infty$ are the $\sigma$-algebras generated by $\{(F_t,e_t):t\leq 0 \}$ and $\{ (F_t,e_t):t\geq T \}$ respectively. \label{assum: mixing}
\end{assumption}

\begin{assumption}
There exists $C > 0$ such that, for all $i \leq p$, $t\leq T$ and $s \leq T$,
\begin{itemize}
\item[(a)] $\left\vert\left\vert b_i\right\vert\right\vert_{max} < C$.
\item[(b)] $\text{E}_{\mathbb{P}^*}\left(p^{-1/2} \left(e_s^{\top} e_t - \text{E}_{\mathbb{P}^*}(e_s^{\top} e_t) \right) \right)^4 < C$, 
\item[(c)] $\text{E}_{\mathbb{P}^*}\left\vert\left\vert p^{-1/2}\sum_{i=1}^p b_i e_{it} \right\vert\right\vert^4 < C$.
\end{itemize}
\label{assum：moment}
\end{assumption}

Assumption~\ref{assum: loading} requires the factors to be pervasive, that is, to impact a non-vanishing proportion of individual asset returns. Assumption~\ref{assum: error term}(a) requires both error term and common factors to be strictly stationary, and $e_t$ and $F_t$ are non-correlated. Assumption~\ref{assum: error term}(b) assumes that $\Sigma_e^*$ is well-conditioned. Assumption~\ref{assum: error term}(c) imposes a sparsity condition on $\Sigma_{e}^{*}$ based on the intuition that many pairs of cross-sectional units become weakly correlated after removing the common factors. In particular, $m_\varsigma$ is the maximum number of non-zero elements in each row when $\varsigma = 0$. Assumption~\ref{assum: error term}(d) imposes exponential-type tails for both errors and common factors. Note that it also implies factor $F_t$ has a finite fourth order moment assumed in Assumption~\ref{assum: clt}(a).
Assumption~\ref{assum: mixing} imposes the strong mixing condition on $F_t$ and $e_t$ with mixing coefficient $\alpha$ decaying exponentially. Also note that the long-run variance-covariance $V_g$ can be consistently estimated by the kernel-based method based on Assumption~\ref{assum: mixing}. In addition, without loss of generality, we assume that  $\text{E}_{\mathbb{P}^*}(F_tF_t^{\top}) = \bm{I}_K$, the similar simplifications in the context of latent factor model that can be found in, for example, \cite{fan2013large} and \cite{li2022integrative}.

Under Assumptions~\ref{assum: loading}-\ref{assum：moment}, the estimated $\widehat{b}_i$ and $\widehat{F}_t$ are consistent with the rotated population factor loading $Hb_i$ and common factors $HF_t$, where rotation matrix $H = \frac{1}{T}\mathcal{V}^{-1}\widehat{F}^\top F B^\top B$ and $\mathcal{V}$ is the $K \times K$ diagonal matrix of the first $K$ largest eigenvalues of the sample covariance of $r_t$, see more details in \cite{fan2013large}. However, the estimator $\widehat{V}_{g, T}$ is established on $\widehat{F}_t$ and thus can not consistently converge to the true one without any further conditions. As a result,  we further provide the identification assumption as follows.
\begin{assumption}
$H = \bm{I}_K + O_p(\Delta)$ where $\Delta/(\sqrt{\log p/T}+T^{1/4}/\sqrt{p}) = o(1)$. \label{assum: identification}
\end{assumption}

Assumption~\ref{assum: identification} requires that the rotation matrix $H$ is approximately equal to an identity matrix. In fact, \cite{bai2013principal} studied conditions under which the latent factors can be estimated asymptotically without rotation. For example, if we impose the identification condition $B^\top B$ is a diagonal matrix with distinct entries, then we can derive that $H = \bm{I}_K + O_{p}(\delta_{pT}^{-2})$ where $\delta_{pT} = \min(\sqrt{p},\sqrt{T})$. A block diagonal matrix of factor loadings satisfies this condition. More examples can be found in \cite{bai2013principal}.
\begin{assumption}
\label{assum: Vg} 
\begin{itemize}
\item[(a)]  The kernel function $k(\cdot)$ has bounded support on $[-1,1]$ and $\left\vert k(x)\right\vert \leq 1$ for all $x$  on the real line, $k(x)=k(-x)$,
$k(0)=1$; $k(x)$ is continuous at zero and for almost all $x$. 
\item[(b)] $q_{T}^{-1} + q_{T}T^{-1/2} + q_{T}T^{1/4} p^{-1/2}\rightarrow 0$ as $T\rightarrow\infty$.
\end{itemize}
\end{assumption}

Assumption~\ref{assum: Vg}(a) gives the requirement of kernel function applied in HAC estimation of $V_g$, which is common,  see e.g., \cite{de2000consistency}. Many classic commonly employed kernel functions, e.g., Bartlett kernel, quadratic spectral kernel and so on, satisfy this Assumption \citep{andrews1991heteroskedasticity, andrews1992improved}. For lag truncation parameter $q_T$, it is required to diverge to infinity, but the rate of divergence is also constrained by the convergence rate of $\widehat{F}_t$ with respect to $F_t$. 
\begin{assumption}\label{assum: rate for portfolio}
\begin{enumerate}
\item[(a)] There exists a constant $M > 0$ such that  $\|V_g\| < M$.
\item[(b)] Suppose the terms $\bm{1}_{p}^{\top}\Sigma_{r}^{-1}\bm{1}_{p} \asymp \bm{1}_p^{\top} \Sigma_{r}^{-1}\mu \asymp \mu^{\top} \Sigma_{r}^{-1}\mu \asymp p^\phi $, and $(\bm{1}_p^\top \Sigma_{r}^{-1}\bm{1}_p)( \mu^\top \Sigma_r^{-1}\mu) - (\bm{1}_p^\top \Sigma_r^{-1}\mu )^2 \asymp p^{2\phi}$.
\item[(c)] $p^{3/2-3\phi}\zeta(p,T) + p^{1-3\phi}m_\varsigma \omega_T^{1-\varsigma} + p^{1-2\phi}q_T\varpi_T = o(1)$, where $\zeta(p,T) =   \sqrt{\frac{logp}{T}} + \frac{T^{1/4}}{\sqrt{p}}$ and $\varpi_T = T^{-1/2} + T^{1/4}p^{-1/2}$. 
\end{enumerate}
\end{assumption}

Assumption~\ref{assum: rate for portfolio}(b) supposes that $\bm{1}_p^\top \Sigma_r^{-1}\bm{1}_{p}, \bm{1}_p^{\top}\Sigma_r^{-1}\mu,$ and $\mu^\top \Sigma_r^{-1}\mu $ have the same order of magnitude. This assumption is made for simplicity; different orders could be assumed for each term. Furthermore,  the assumption that the term $\bm{1}_p^\top \Sigma_r^{-1}\bm{1}_{p}$ is of the order of powers of $p$ can also be found in \cite{ding2021high} and \cite{fan2022}. Assumption~\ref{assum: rate for portfolio}(b)  indicates that the assets in the investing pool are well diversified such that $\mu$ is not close to $C_1\bm{1}_p$ for some constant $C_1$. Assumption~\ref{assum: rate for portfolio}(c) guarantees that  $\widehat{w}_{mv}$ converges to $w_{mv}^{*}$ under the max norm. 

\subsection{Asymptotic Consistency}

The following Theorems~\ref{Theorem: mufmufhat -muf muf} and~\ref{theorem Vg consistent} provide the consistency of $\widehat{\mu}_f^\top \widehat{\mu}_f$ and $\widehat{V}_{g,T}$ to their population counterpart, respectively:

\begin{theorem}\label{Theorem: mufmufhat -muf muf}
Suppose Assumptions~\ref{assum: clt}, ~\ref{assum: loading} --~\ref{assum：moment} hold true, $y^{-1}= 3y_1^{-1}+1.5y_2^{-1}+y_3^{-3}+1$,  $log p = o(T^{y/6})$, and $T = o(p^2)$. Then as $T\rightarrow \infty$, we have $\widehat{\mu}_f^\top \widehat{\tilde{\Sigma}}_f^{-1}\widehat{\mu}_f \overset{p}{\rightarrow} \mu_f^{*\top}\tilde{\Sigma}_f^{*-1}\mu_f^{*}$.
\end{theorem}

\begin{theorem}\label{theorem Vg consistent}
Under the conditions of Theorem~\ref{Theorem: mufmufhat -muf muf},  and Assumptions~\ref{assum: identification} and~\ref{assum: Vg} hold true,  as $T\rightarrow \infty$, we have $\widehat{V}_{g,T}\overset{p}{\rightarrow} V_{g}$.
\end{theorem}



If $F_t$ is independent across time horizon, $V_g$ degenerates to the covariance matrix of $F_t$, and it is straightforward to obtain that $\widehat{V}_{g, T} = \bm{I}_K -\widehat{\mu}_f\widehat{\mu}_f^\top$ which is a consistent estimator according to Theorem~\ref{Theorem: mufmufhat -muf muf}.  Based on Theorem~\ref{Theorem: mufmufhat -muf muf} and~\ref{theorem Vg consistent}, we can consistently estimate the 
$1-\alpha_0$ quantile of $L_0 = {\| Z_0\|_2^2}/{4\left( 1- \mu_f^{*\top}\tilde{\Sigma}_f^{*-1}\mu_f^*\right)}$, and further consistently estimate $\delta$ chosen from Section~\ref{sec: choice of delta}.  Following Theorem~\ref{Theorem: wBVBw consistent}  provides the consistency of estimated variance of normal variate $Z_1$ in~\eqref{eq: clt for rho}.
\begin{theorem}\label{Theorem: wBVBw consistent}
Under the conditions of Theorem~\ref{theorem Vg consistent}, and Assumption \ref{assum: rate for portfolio} holds true, then we have $\widehat{w}_{mv}^{\top}\widehat{B}\widehat{V}_{g,T}\widehat{B}^{\top}\widehat{w}_{mv} \overset{p}{\rightarrow} w^{*\top}_{mv}B V_{g} B^{\top} w^{*}_{mv}$.
\end{theorem}

As an immediate result of Theorem~\ref{Theorem: wBVBw consistent}, we know that $\|\widehat{B}^\top \widehat{w}_{mv}\|_2$ is consistent with $\|B^\top w_{mv}^*\|_2$ by letting $\widehat{V}_{g,T}= V_g = I_K$. Therefore, up to this point,  the value of $\rho$ selected from Section~\ref{sec: choice of rho} can be consistently estimated when $q = 2$. It is worth noting that the consistency of $\|\widehat{B}^\top \widehat{w}_{mv}\|$ does not require the identification condition \ref{assum: identification}.


\section{Simulation}\label{sec:6}

In this section, we conduct Monte Carlo simulations to evaluate the performance of the selection procedure of $\delta$ and $\rho$ as well as our distributionally robust portfolio.

\subsection{Basic Set-up}\label{sec: basic setup}

The virtual return data are generated by a factor model defined in~\eqref{factor model} with two common factors. We impose an AR$(1)$ structure to each factor, that is 
\begin{equation*}
f_{i,t} = \beta_i + \alpha_i f_{i,t-1} + v_{t}, 
\end{equation*}
where $v_{t} \sim N\left(0, (1-\alpha_{i}^2)(1-\beta_i^2/(1-\alpha_i)^2 \right)$ such that $\text{E}_{\mathbb{P}^*}(f_{i,t}^2) = 1$. The AR(1) specification for latent common factors is often applied in factor model-based financial literature; see e.g., \cite{su2017time} and \cite{fan2022}. For factor loading $B$, we set it to  a block diagonal matrix 
\begin{equation*}
B = \left(\begin{array}{cc}
B_1  & \textbf{0}_{p_1} \\
\textbf{0}_{p_2} & B_2 
\end{array} \right), 
\end{equation*}
where $B_i$ is a vector of $p_i$ with $p_1 + p_2 = p$, and $B_i \sim N(\mu_{b_i},\sigma_{b_i}^2)$. For error term $e_t$, we simply draw them from normal distributions  $N(0,\Sigma_e)$. The values of population parameters are calibrated from real daily return data of the largest $p$ components of the S\&P 500 index (measured by market values)  from 01/2006 to 12/2009. In more detail, we conduct principal component analysis to estimate the latent factors, factor loadings, and corresponding residuals. Parameters $\alpha_i$ and $\beta_i$ are set to the autocoefficient and intercept estimations fitted by the AR(1) process using estimated factors, and $\mu_{b_i}$ and $\sigma_{b_i}^2$ are given by the mean and variance of estimated loading. For $\Sigma_e^*$, we assign the estimated sparse error covariance matrix (\ref{eq: estimated sparse cov}) based on estimated residuals with threshold $\tau_{ij} = 0.5\sqrt{\widehat{{\theta}}_{ij}}\left(\sqrt{\log p/T} + 1/\sqrt{p} \right)$ where $\widehat{{\theta}} = \frac{1}{T}\sum_{t=1}^T\left(\widehat{e}_{it}\widehat{e}_{jt} - s_{ij} \right)^2$.

\subsection{Performance of Selected \texorpdfstring{$\delta$}{delta}}\label{sec: simu delta}

A good selection for $\delta$ is important for distributionally robust portfolio allocation. The choice of $\delta$ in BCZ is criticized as too large, which leads to an unnecessarily conservative portfolio. In our set-up, even though $\delta$ is chosen as too large, the information of factor loading and error covariance matrix also contribute to the final portfolio weights, thus the optimal allocation is unlikely to be an equal weighing allocation. Regardless, in this subsection, we evaluate the estimated value of $\delta$ via simulation. The results imply that the $\delta$ chosen from our procedure is quite close to the true level.   For other set-ups, the target return $\bar{\alpha}$  of problem~\eqref{model:classic} is set to 0.05\%, considering that the values of all population parameters are calibrated from daily return data. Note that the 0.05\% daily excess return corresponds to 13.42\% annualized excess return. We note that only the procedure developed by BCZ requires the value of the target portfolio return when choosing $\delta$. For notational simplicity, the selected $\delta$ based on population parameter and estimated parameter are denoted as $\delta_{hd}$ and $\widehat{\delta}_{hd}$, respectively. We generate 500 virtual dataset $R_{p\times T}$ with $T= 200$ and $p \in \{30,50,80,100\}$. For estimator $\widehat{V}_{g,T}$,  we choose Bartlett kernel function $k(x) = (1-|x|)I(x\leq 1)$ which can be verified to satisfy Assumption \ref{assum: Vg}, see \cite{andrews1991heteroskedasticity}, and  bandwidth {$q_T = \lfloor cT^{-1/8}p^{1/4} \rfloor$ } with constant $c = 5$ given  that $q_TT^{1/4}p^{-1/2} = o(1)$. For comparison, we report the true and estimated $\delta$ values from BCZ, and denote them as $\delta_{bl}$and $\widehat{\delta}_{bl}$, respectively. 

Table~\ref{Table delta simulation} reports the median and mean level of estimated  $\widehat{\delta}$ over 500 replications as well as the corresponding oracle values. The values in brackets are the ratio of estimated values over the corresponding population, indicating the deviation from the true level. It is evident that the median and mean values of $\widehat{\delta}_{bl}$ largely deviate from their true values when $p/T$ is large. However, those of $\widehat{\delta}_{hd}$ are close to the true values in all cases.  For example, in the case of $p= 50$, the median of $\widehat{\delta}_{bl}$ at the 95\%  confidence level is 8.62E-04, which is about 14 times the oracle value, whereas the value of $\widehat{\delta}_{hd}$ is approximately equal to the corresponding oracle value. We note that the comparison based on absolute values is meaningless since the values of $F_t$ and $r_t$ are incomparable due to the existence of factor loading. Furthermore,
the deviations of $\widehat{\delta}_{bl}$ from the true values increase substantially as $p$ gets larger.  Additionally, one can also observe that there is a large gap between the mean and median values of $\widehat{\delta}_{bl}$, indicating that there exist many extremely poor estimations over simulation replications. In contrast, the mean and median values of $\widehat{\delta}_{hd}$ are very close, which indicates the robustness of $\widehat{\delta}_{hd}$. 

\begin{table}[!htb]
\centering
\tabcolsep 0.2in
\caption{The median and mean level of $\widehat{\delta}$ from BCZ (denoted as $\widehat{\delta}_{bl}$) and our proposal (denoted as $\widehat{\delta}_{hd}$), and corresponding oracle values. The ratio compared to the corresponding population value is reported in brackets. The sample size of the training set is 200, and the target return $\overline{\alpha}$ for \eqref{model:classic}  is  0.0005. The 90\%, 95\% and 99\% confidence levels for choosing $\delta$ are considered.}\label{Table delta simulation}
\begin{tabular}{@{}lccccccc@{}}
\toprule
                     & \multicolumn{3}{c}{$\delta$ from BCZ} &                      & \multicolumn{3}{c}{Our $\delta$} \\ \cline{2-4} \cline{6-8} 
$p$                    & 90\%      & 95\%      & 99\%     &                      & 90\%              & 95\%              & 99\%              \\ \midrule
                     & \multicolumn{7}{c}{Oracle}                                                                                          \\
30                   & 3.50E-05  & 4.12E-05  & 5.61E-05 & \multicolumn{1}{l}{} & 6.36E-03          & 8.31E-03          & 1.30E-02          \\
50                   & 5.35E-05  & 6.23E-05  & 8.36E-05 & \multicolumn{1}{l}{} & 5.86E-03          & 7.65E-03          & 1.19E-02          \\
80                   & 8.81E-05  & 1.03E-04  & 1.39E-04 & \multicolumn{1}{l}{} & 5.67E-03          & 7.40E-03          & 1.15E-02          \\
100                  & 1.23E-04  & 1.48E-04  & 2.10E-04 &                      & 5.59E-03          & 7.30E-03          & 1.12E-02          \\ \midrule
& \multicolumn{7}{c}{Median}                                                                                          \\
30                   & 2.69E-04  & 3.17E-04  & 4.30E-04 &                      & 6.11E-03          & 7.98E-03          & 1.24E-02          \\
                     & (7.68)      & (7.70)      & (7.66)     &                      & (0.96)              & (0.96)              & (0.96)              \\
50                   & 7.31E-04  & 8.62E-04  & 1.19E-03 &                      & 5.74E-03          & 7.56E-03          & 1.18E-02          \\
                     & (13.66)    & (13.84)    & (14.22)   &                      & (0.98 )             & (0.99)            & (0.99)            \\
80                   & 2.17E-03  & 2.57E-03  & 3.54E-03 &                      & 5.72E-03          & 7.48E-03          & 1.16E-02          \\
                     & (24.60)    & (24.97)     & (25.53)   &                      & (1.01)              & (1.01)            & (1.01)            \\
100                  & 3.94E-03  & 4.79E-03  & 6.81E-03 &                      & 5.69E-03          & 7.40E-03          & 1.15E-02          \\
                     & (32.14)    & (32.42)    & (32.41)   &                      & (1.02)             & (1.01)             & (1.02)             \\ \midrule
 & \multicolumn{7}{c}{Mean}                                                                                            \\
30                   & 7.09E-02  & 8.39E-02  & 1.15E-01 &                      & 6.15E-03          & 8.05E-03          & 1.25E-02          \\
                     & (2022.86)  & (2036.38)  & (2043.23)  &                      & (0.97)              & (0.97)              & (0.96)             \\
50                   & 1.77E-03  & 2.11E-03  & 2.93E-03 &                      & 5.82E-03          & 7.62E-03          & 1.19E-02          \\
                     & (33.09)    & (33.85)    & (35.11)   &                      & (0.99)             & (1.00)             & (1.00)             \\
80                   & 2.84E-03  & 3.37E-03  & 4.69E-03 &                      & 5.77E-03          & 7.55E-03          & 1.18E-02          \\
                     & (32.20)    & (32.79)    & (33.80)   &                      & (1.02)              & (1.02)             & (1.02)              \\
100                  & 4.52E-03  & 5.46E-03  & 7.79E-03 &                      & 5.71E-03          & 7.49E-03          & 1.17E-02          \\
\multicolumn{1}{l}{} & (36.88)     & (36.98)     & (37.09)    & \multicolumn{1}{l}{} & (1.02)             & (1.03)             & (1.04)  \\
\bottomrule
\end{tabular}
\end{table}

To see more insights into estimated values of $\delta$, Table \ref{Table delta simulation more detail} further presents the standard deviation, the ratio of standard deviation to median level, maximum value, and the ratio of maximum value to median level for both $\widehat{\delta}_{cl}$ and $\widehat{\delta}_{hd}$ over 500 simulation replications. First, it can be observed that the standard deviations of $\widehat{\delta}_{bl}$ are quite large compared to its median level. For example, in the case of $p=30$, the standard deviation at three confidence levels is  3,000 times the corresponding median level.  In contrast, the volatility of $\widehat{\delta}_{hd}$ looks to be normal, the value is only 10\% of the corresponding median level. Similarly, the maximum value of $\widehat{\delta}_{bl}$ over simulation is far away from the median level. For example, the maximum value is 8 times the median level in the best performance case of $p = 100$, while the maximum value of  $\widehat{\delta}_{hd}$ is  close to only 1.5 times the median level. It is worth noting that even though the multiples of the standard deviation (maximum values) decrease compared to the median as $p/T$ gets larger, the median level also deteriorates quickly. Clearly, $\widehat{\delta}_{bl}$ is an unstable estimation, which is detrimental to a method known for its robustness, and hence our proposal is favored in practical application.
\begin{table}[!htb]
\centering
\tabcolsep 0.16in
\caption{The standard deviation, the ratio of standard deviation to median level, maximum value and the ratio of maximum value to median level for both $\widehat{\delta}_{cl}$ and $\widehat{\delta}_{hd}$ over simulation replications. The sample size of the training set is 200, and the target return $\overline{\alpha}$ for \eqref{model:classic}  is  0.0005. The 90\%, 95\% and 99\% confidence levels for choosing $\delta$ are considered. }
\label{Table delta simulation more detail}
\begin{tabular}{@{}llrrrrrrr@{}}
\toprule
& \multicolumn{1}{l}{} & \multicolumn{3}{c}{$\delta$ from BCZ} & \multicolumn{1}{c}{} & \multicolumn{3}{c}{Our $\delta$} \\ \cline{3-5} \cline{7-9} 
$p$   & \multicolumn{1}{l}{} & 90\%              & 95\%              & 99\%              & \multicolumn{1}{c}{} & 90\%      & 95\%      & 99\%     \\ \hline
30  & SD                   & 0.898             & 1.058             & 1.450             &                      & 0.001     & 0.001     & 0.002    \\
 & SD/Median            & 3335.906          & 3333.615          & 3374.706          &                      & 0.118     & 0.120     & 0.134    \\
                        & Max                  & 14.694            & 17.324            & 23.687            &                      & 0.009     & 0.011     & 0.019    \\
                        & Max/Median           & 54587.347         & 54564.406         & 55118.921         &                      & 1.444     & 1.402     & 1.499    \\ \cmidrule{2-9}
50  & SD                   & 0.008             & 0.010             & 0.014             &                      & 0.001     & 0.001     & 0.002    \\
   & SD/Median            & 11.401            & 11.572            & 11.729            &                      & 0.123     & 0.123     & 0.136    \\
                        & Max                  & 0.158             & 0.190             & 0.267             &                      & 0.009     & 0.012     & 0.018    \\
                        & Max/Median           & 215.646           & 220.026           & 225.004           &                      & 1.631     & 1.618     & 1.502    \\ \cmidrule{2-9}
80  & SD                   & 0.003             & 0.003             & 0.005             &                      & 0.001     & 0.001     & 0.002    \\
    & SD/Median            & 1.255             & 1.254             & 1.295             &                      & 0.127     & 0.129     & 0.140    \\
                        & Max                  & 0.037             & 0.044             & 0.064             &                      & 0.009     & 0.011     & 0.018    \\
                        & Max/Median           & 17.216            & 17.016            & 18.071            &                      & 1.522     & 1.531     & 1.579    \\ \cmidrule{2-9}
100 & SD                   & 0.003             & 0.004             & 0.005             &                      & 0.001     & 0.001     & 0.002    \\
                        & SD/Median            & 0.737             & 0.736             & 0.759             &                      & 0.121     & 0.126     & 0.138    \\
                        & Max                  & 0.033             & 0.041             & 0.056             &                      & 0.008     & 0.011     & 0.018    \\
                        & Max/Median           & 8.401             & 8.539             & 8.180             &                      & 1.435     & 1.488     & 1.561   \\
\bottomrule
\end{tabular}
\end{table}

\begin{figure}[!htb]
\centering	
\includegraphics[width=9cm, trim={0cm 0 0 0}]{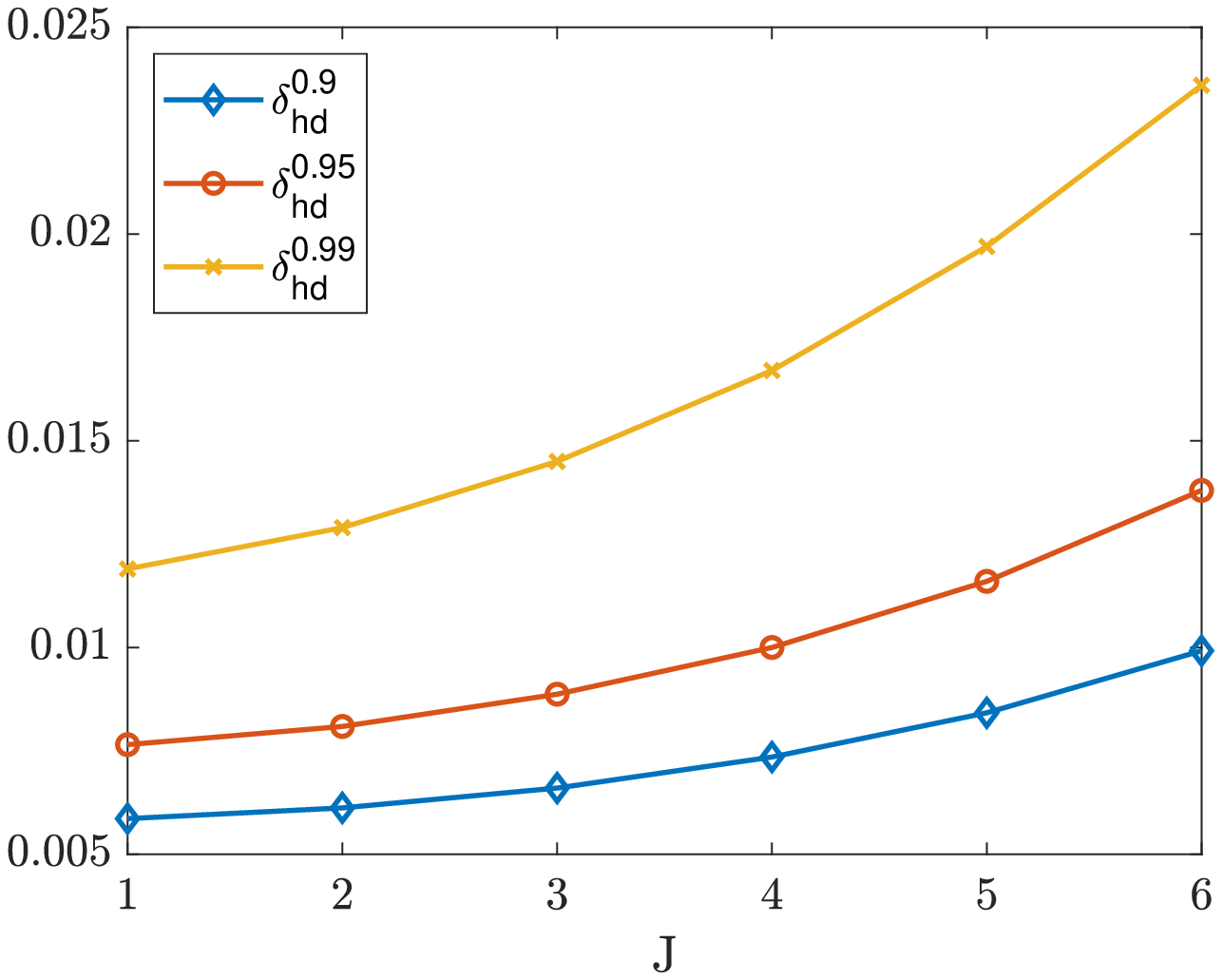}	
\caption{The model uncertainty measured by $\delta_{hd}$ under confidence levels $0.9, 0.95$, and $0.99$.}\label{Figure simulation measure uncertainty}
\end{figure}

For a distributionally robust approach, it is interesting to study the relationship between model uncertainty and the selected $\delta_{hd}$. It is expected that the value of $\delta_{hd}$ increases as model uncertainty increases.
For a factor with an AR(1) structure, an easy way to alter the uncertainty is to change the autocoefficient $\alpha_i$. When $\alpha_i$ is close to zero, the factor tends to be independent across the time horizon. When $\alpha_i$  is close to 1, the factor has a long persistent correlation. Clearly, long-term correlation will result in more uncertainty than independence.  Let $\alpha_i = J\alpha_{i0}$ where $\alpha_{i0}$ is the value calibrated from real data, and the previous results are based on the situation of $J = 1$. Figure \ref{Figure simulation measure uncertainty} plots uncertainty measures $\delta_{hd}$ with respect to the cases of different $J\in \{1,2,\ldots,6\}$ at the 0.9, 0.95 and 0.99 confidence levels\footnote{When $J =6$, two factors are still stationary.}. 
It is evident that  $\delta_{hd}$ takes larger value as $J$ gets larger, indicating that the model uncertainty  increases as $J$ gets larger. Thus, we can conclude that the value of $\delta$ from our selection procedure well reflects the size of model uncertainty.


\subsection{Performance of Selected \texorpdfstring{$\rho$}{alpha}}\label{sec: simu rho}

In this subsection, we further study the performance of choosing $\rho$. Recall that the choice of $\rho$ is given by $\overline{\alpha} - \sqrt{\delta}\|B^\top w_{mv}^*\|_2 v_{0}$. It is clear that the choice of $\delta$ would also affect the choice of $\rho$.  Notice that the key step to selecting the value for $\rho$ is the central limit theorem shown in \eqref{eq: clt for rho}. Thus, to eliminate the impact of choosing $\delta$, we focus on estimating the quantity $Q = T^{-1/2}\mathcal{A}/\| B^\top w_{mv}^*\|_2$. Once $Q$ is obtained, the value of  $\sqrt{\delta}(1-v_0)$ is determined and thus the choice of $\rho$ is also determined.  We denote the estimated (oracle) choices from our work and that of BCZ by $\widehat{Q}_{hd}$ ($Q_{hd}$) and $\widehat{Q}_{bl}$ ($Q_{bl}$), respectively. The simulation set-ups remain unchanged from those used in the previous subsection. 

Table~\ref{Table delta simulation rho} reports the median and mean level of $\widehat{Q}_{hd}$ and $\widehat{Q}_{bl}$ over 500 replications as well as the corresponding oracle values. We consider 90\%, 95\%, and 99\% confidence levels. It is evident that $\widehat{Q}_{bl}$ is severely underestimated compared to the true level, while $\widehat{Q}_{hd}$ is quite close to the oracle level. For example, in the case of $p = 100$,  at all three confidence levels,  $\widehat{Q}_{bl}$ is only 18\% of the true value, whereas $\widehat{Q}_{hd}$ is 99\% of corresponding true value. Fortunately, unlike the $\widehat{\delta}_{bl}$, the estimation $\widehat{Q}_{bl}$ is much more stable, which is indicated by the closeness of mean and median values. One can also observe that as $p$ gets larger, estimator $Q_{hd}$ also gets closer to the oracle level.

\begin{table}[!htb]
\centering
\tabcolsep 0.19in
\caption{The median and mean level of $\widehat{Q}_{hd}$ and $\widehat{Q}_{bl}$, and corresponding oracle levels. The ratio compared to the corresponding oracle level is reported in brackets. The sample size of the training set is 200, the target return $\overline{\alpha}$ in \eqref{model:classic}  is  0.05\%. The 90\%, 95\% and 99\% confidence levels are considered. }
\label{Table delta simulation rho}
\begin{tabular}{@{}lccccccc@{}}
\toprule
                     & \multicolumn{3}{c}{BCZ} &                      & \multicolumn{3}{c}{Our proposal}       \\ \cline{2-4} \cline{6-8} 
$p$                    & 99\%             & 95\%            & 90\%            &                      & 99\%      & 95\%      & 90\%      \\ \midrule
                     & \multicolumn{7}{c}{Oracle}                                                                                      \\
30                   & -4.87E-03 & -3.44E-03 & -2.68E-03 & \multicolumn{1}{l}{} & -1.85E-01 & -1.31E-01 & -1.02E-01 \\
50                   & -7.26E-03 & -5.13E-03 & -4.00E-03 & \multicolumn{1}{l}{} & -1.75E-01 & -1.24E-01 & -9.64E-02 \\
80                   & -9.74E-03 & -6.89E-03 & -5.36E-03 & \multicolumn{1}{l}{} & -1.69E-01 & -1.19E-01 & -9.30E-02 \\
100                  & -8.01E-03 & -5.66E-03 & -4.41E-03 & \multicolumn{1}{l}{} & -1.68E-01 & -1.19E-01 & -9.24E-02 \\ \midrule
                     & \multicolumn{7}{c}{Median}                                                                                      \\
30                   & -2.10E-03 & -1.48E-03 & -1.16E-03 &                      & -1.70E-01 & -1.21E-01 & -9.39E-02 \\
                     & (0.43)      & (0.43)      & (0.43)      &                      & (0.92)      & (0.92)      & (0.92)      \\
50                   & -1.99E-03 & -1.41E-03 & -1.10E-03 & \multicolumn{1}{l}{} & -1.66E-01 & -1.17E-01 & -9.12E-02 \\
                     & (0.27)      & (0.27)     & (0.27)      &                      & (0.95)      & (0.95)      & (0.95)      \\
80                   & -1.68E-03 & -1.19E-03 & -9.28E-04 &                      & -1.66E-01 & -1.17E-01 & -9.12E-02 \\
                     & (0.17)      & (0.17)      & (0.17)      &                      & (0.98)      & (0.98)      & (0.98)      \\
100                  & -1.47E-03 & -1.04E-03 & -8.09E-04 &                      & -1.65E-01 & -1.17E-01 & -9.11E-02 \\
                     & (0.18)      & (0.18)      & (0.18)      &                      & (0.99)      & (0.99)      & (0.99)   \\ \midrule
\multicolumn{1}{l}{} & \multicolumn{7}{c}{Mean}                                                                                        \\
30                   & -2.12E-03 & -1.50E-03 & -1.17E-03 &                      & -1.71E-01 & -1.21E-01 & -9.40E-02 \\
                     & (0.44)      & (0.44)      & (0.44)      &                      & (0.92)      & (0.92)      & (0.92)      \\
50                   & -2.01E-03 & -1.42E-03 & -1.11E-03 &                      & -1.66E-01 & -1.17E-01 & -9.12E-02 \\
                     & (0.28)      & (0.28)     & (0.28)      &                      & (0.98)      & (0.98)      & (0.98)      \\
80                   & -1.69E-03 & -1.20E-03 & -9.33E-04 &                      & -1.65E-01 & -1.17E-01 & -9.12E-02 \\
                     & (0.17)      & (0.17)      & (0.17)      &                      & (0.98)      & (0.98)      & (0.98)     \\
100                  & -1.48E-03 & -1.04E-03 & -8.14E-04 &                      & -1.65E-01 & -1.17E-01 & -9.11E-02 \\
\multicolumn{1}{l}{} & (0.18)      & (0.18)      & (0.18)      & \multicolumn{1}{l}{} & (0.99)      & (0.99)      & (0.99)   \\
\bottomrule
\end{tabular}
\end{table}

\subsection{Performance of Portfolio}\label{sec: simu port}

In the previous two subsections, we have demonstrated that the estimating procedure in Algorithm \ref{algonew delta} can obtain accurate and stable $\widehat{\delta}_{hd}$ and $\widehat{\rho}$. In this subsection, we further study the performance of a high dimensional distributionally robust portfolio obtained from solving problem \eqref{our model reform}.

In Section \ref{sec:2}, we noted that  the target portfolio return constraint  $w^\top B \text{E}_{\mathbb{P}_T}(F_t) \geq \rho + \sqrt{\delta}||B^\top w||_2$ in \eqref{our model reform} may lead to an empty feasible region given $\delta, B$, and $\text{E}_{\mathbb{P}_T}(F_t)$.  To better see this, let us define  $\mathcal{G} =  w^\top \mu^* - \sqrt{\delta}\| B^\top w \|_2  $ where $\mu^* = B\mu_f^*$, and  $\underline{\mathcal{G}} = min_w \mathcal{G} $ with respect to the budget constraint $w^\top \textbf{1}_p = 1$. Then we know that the quantity $\underline{\mathcal{G}} $ is the largest allowable $\rho$ for which the problem \eqref{our model reform} has a non-empty feasible region. Under the basic set-up defined in Subsection \ref{sec: basic setup}, and let $\delta$  be  $\delta_{hd}$ calculated at the $95\%$ confidence level, the value of  $\underline{\mathcal{G}} $ for the cases under various dimensions are reported in Table \ref{Table: maximum rho}. We can observe that the values of $\underline{\mathcal{G}}$ are quite close to $0$ from the negative direction. It implies that if we choose a positive value for $\rho$, we cannot obtain the distributionnally robust portfolio via problem \eqref{our basic model}. Further note that the maximum allowable $\rho$ for problem \eqref{blenchet2022 model2} is about $-0.001$ under the same set-up, that is, it has a stricter requirement for $\rho$ compared to our model. As a result, in practical application, the value of $\rho$ cannot be chosen randomly based on user's preferences and is crucial for obtaining a distributionally robust portfolio allocation. Fortunately, the choice of $\rho$ from the procedure in Subsection \ref{sec: choice of rho} can be applied directly.
Recall that, the value of $\rho$ is selected such that we have the inclusion $w^*_{mv} \in \mathcal{F}_{\delta,\rho}(T)$ with a given  confidence level $1-\epsilon$, where  $\mathcal{F}_{\delta,\rho}(T)$ is the feasible region. It indicates that we are $1-\epsilon$ confident that the feasible set contains at least one feasible point $w_{mv}^*$, and thus the feasible region using selected $\rho$ wouldn't be empty.

Table~\ref{Table: maximum rho} also reports the true choice  $\rho_{hd}$ for~\eqref{our model reform}. The 95\% confidence level for both $\delta$ and $\rho$ are applied. We can observe that the selected values of $\rho$ are smaller than the corresponding upper bound, which indicates that there exist some portfolio allocations satisfying the constraints in problem \eqref{our model reform}.

\begin{table}[!htb]
\centering
\tabcolsep 0.5in
\caption{The values of $\overline{\mathcal{G}}$ and $\rho_{hd}$  under the basic set-up.}\label{Table: maximum rho}
\begin{tabular}{@{}lcr@{}}
\toprule
$p$ &  $\overline{\mathcal{G}}$ &  ${\rho}_{hd}$ \\
\midrule
30  &  -5.53E-10                   & -3.50E-03                   \\
50  &  -5.27E-10                   &  -2.92E-03                  \\
80  &  -7.29E-10                   &  -3.58E-03                  \\
100 &  -1.12E-09                   &  -3.26E-03                 \\ 
\bottomrule
\end{tabular}
\end{table}

In order to conduct an  out-of-sample performance evaluation, we generate 500 virtual datasets $R_{p\times 2T} = [R_{1,p\times T}, R_{2,p\times T}]$ with $T = 200$. The first 200 periods of data $R_{1,p\times T}$ are used as the training set to construct various considered portfolios, and the last 200 periods of data $R_{2,p\times T}$ are treated as the testing set. We denote the portfolio from the proposed procedure by ``HD-DRO”, and the portfolio from \cite{blanchet2022distributionally} by ``DRO". In addition, the results of equal weighting strategy (1/N) are also reported for comparison.   The confidence level for $\delta$ and $\rho$ are both set to 95\%.   The out-of-sample standard deviation is reported and is defined as the standard deviation of realized out-of-sample portfolio excess returns. 
We note that the risk measure is important for evaluating the robust portfolio, since investors using this approach are more concerned about the risk they undertake and the lowest acceptable portfolio return $\rho$ being less than 0 indicates that the strategy does not aim at high return.

Table~\ref{Table simulation port SD} reports the out-of-sample standard deviation of HD-DRO, DRO, and 1/N for the cases of $p \in \{30,50,80,100, 150\}$. The values in the bracket are the ratio compared to the corresponding oracle results. First, it can be observed that the HD-DRO achieves the lowest out-of-sample standard deviation in all cases. The risk of HD-DRO decreases as $p$ gets larger. For DRO, the risks are larger than those of HD-DRO but smaller than those of the 1/N strategy. Furthermore, compared to the corresponding oracle version, the performance of HD-DRO is closer to its population counterpart than that of DRO. For example, the ratio of HD-DRO in the case of $p = 80$ is 1.09, whereas that of DRO is 1.20. For DRO, the deviation from the oracle gets larger as $p$ increases, whereas the deviation for HD-DRO to its true level is much more stable. We note that even though the simulation results in Table~\ref{Table delta simulation more detail} shows the obvious upward biases of $\widehat{\delta}_{bl}$ compared to the true values, the performances of DRO are still far away from those of  1/N strategy on average level. It seems that the portfolio allocation is less affected by the overestimated regularization parameter than expected. However, we emphasize that, in 500 simulation experiments, there are still some situations where the weights of DRO are very close to equal weights.

\begin{table}[!htb]
\centering
\tabcolsep 0.4in
\caption{The out-of-sample standard deviation of HD-DRO, DRO, and 1/N. The values in the bracket are the ratios compared to the corresponding oracle results. The values of $\delta$ and $\rho$ are chosen by the corresponding procedure at the confidence level of 0.95. The target return in \eqref{model:classic} is 0.0005.}\label{Table simulation port SD}
\begin{tabular}{@{}lccc@{}}
\toprule
$p$ & HD-DRO & DRO & 1/N \\
\midrule
30  & 0.00453 & 0.00509 & 0.00832 \\
& (1.040) & (1.141) &
\\ 
50  & 0.00324 & 0.00359 & 0.00780 \\
& (1.060) & (1.156) &
\\80  & 0.00272 & 0.00306 & 0.00800 
\\
& (1.089) & (1.196) &
\\
100 & 0.00252 & 0.00295 & 0.00863 \\
& (1.082) & (1.229) &
\\
150 &0.00231 &0.00314 &0.00870
\\
& (1.096) & (1.434) &
\\
\bottomrule
\end{tabular}
\end{table}


Finally, we also simulated the cases where the lowest acceptable portfolio return $\rho$ is fixed at some negative values, e.g., $-0.002$. The results are quite similar to those presented in Table~\ref{Table simulation port SD}, thus we do not repeat the illustration of the results.



\section{Empirical Analysis}\label{sec: empirical}

In this section, we evaluate the out-of-sample performance of our proposed high-dimensional distributionally robust portfolio using real market data. Specifically, we construct portfolios based on a rolling-window scheme. At each decision point $t_c$, we use the historical excess return data of length T (before $t_c$) to construct portfolios, which are then held for the next $T_h$ periods to gain the realized excess return. After holding $T_h$ periods, we re-calculate and re-balance the portfolio weights using historical data at the time point $t_c + T_h + 1$. This process is repeated until the end of the out-of-sample period.

\subsection{Choice of Common Factor Number}
In practical implementation, the number of common factors is unknown and needs to be estimated, thus we apply the standard and commonly used approach proposed by \cite{bai2002determining} to determine the number of common factors $K$. In detail, we solve the problem as follows:
\begin{align*}
\widehat{K} = \mathop{\text{arg min}}_{0 \leq K_1 \leq M } \log\left(\frac{1}{pT}\left\vert\left\vert R - T^{-1}R\widehat{F}(K_1)\widehat{F}(K_1)^\top \right\vert\right\vert^2_F \right) + K_1g(T,p) 
\end{align*}
where $M$ is a prescribed upper bound for the number of factors, $\widehat{F}(K_1)$ is estimated common factor conditional on factor number $K_1$, and $g(T,p_m)$ is a penalty function of $p$ and $T$, for example, $g(T,p) = \frac{p+T}{pT}\log(\frac{pT}{p+T})$. The conditions assumed in \cite{bai2002determining} can be guaranteed by our Assumptions. As a result, the consistency of $\widehat{K}$ to $K$ is ensured by their Theorem~2. 
 
\subsection{Choice of Threshold Parameter}\label{sec: choice of tau}

Recall that the thresholding parameter is defined as $\tau_{ij} = C\sqrt{\widehat{\theta}_{ij}}\omega_T$. In implementation, we also need to choose the value for $C$. In more detail, we use the data-driven $m$-fold  cross-validation method to determine it. After we obtain the estimation of error term $\widehat{e}_t$, we randomly divide the residual sequence into two subsets, denoted as $\{\widehat{e}_t\}_{t\in A} $ and $\{\widehat{e}_t\}_{t\in B}$ respectively. Subset $A$ is used for training and subset $B$ is used for testing purposes. Then, we can choose the value of  $C$ that minimizes the following objective function over a compact interval:
\begin{align*}
C^{*} = \mathop{\text{argmin}}_{\underline{C}<C_1 \leq \overline{C}} \frac{1}{m}\sum_{j=1}^{m} \left\|\widehat{\Sigma}_e^{A,j}(C_1) - S_e^{B,j} \right\|^2_F,
\end{align*}
where $\underline{C}$ is the minimum constant that guarantees the positive definiteness of $\widehat{\Sigma}_e^{A,j}(C_1)$, and $\overline{C}$ is large constant such that $\widehat{\Sigma}_e^{A,j}(C_1)$ is diagonal, $\widehat{\Sigma}_e^{A,j}(C)$ is the threshold residual covariance estimator by using subset $A$ in $j$\textsuperscript{th} loop with threshold $C_1$, and $S_e^{B,j} $ is sample covariance matrix by using subset $B$ in $j$\textsuperscript{th} loop. We set the training set to include 2/3 of the total observations.

\subsection{Measures and Benchmark}

The performance of the portfolio is evaluated by its out-of-sample cumulative excess return, risk, Sharpe ratio, and maximum drawdown, which are defined as follows respectively:
\begin{align*}
\text{CR} &= \sum_{t=1}^{T} r_{t}^{p},    \\ 
\text{Risk} &= \sqrt{\frac{1}{T-1}\sum_{t=1}^T\left(r_t^p - \frac{\text{CR}}{T}\right)^2}, \\ 
\text{SR} &= \frac{\text{CR}}{T\times \text{Risk}}, \\ 
\text{MDD} &= \max_{1\leq t_1 \leq t_2 \leq T}-\sum_{t= t_1}^{t_2}r_{t}^{p},
\end{align*}
where $T$ is the number of out of sample observations and $r_t^p$ is out of sample portfolio return.

For competing investment strategies, the results of the distributionally-robust portfolio (\textbf{DRO}) of BCZ and the equal weight strategy (\textbf{1/N})  \citep{demiguel2009optimal} are reported as the main benchmarks.
We follow the menu at the end of Section 4 of BCZ to choose the tuning parameters. BCZ's empirical analysis indicates that the DRO demonstrates its superiority over other robust models, such as \cite{goldfarb2003robust}. For better comparison and simplicity, we do not focus much on the models that have already been considered by BCZ. Additionally, we compare our results with the mean-variance portfolio based on the POET covariance and mean estimators (\textbf{MV-POET}) proposed by \cite{fan2013large}, as we apply  factor-based covariance matrix estimator. We also compare our results with the regularized mean-variance portfolio (\textbf{MV-L1})  with $l_1$ norm developed by \cite{demiguel2009generalized} and the classic mean-variance portfolio (\textbf{MV}) based on sample mean and sample covariance. Our strategy is simply labeled as \textbf{HD-DRO}. The threshold parameter used in our approach and the POET estimator is chosen by the procedure introduced in Section~\ref{sec: choice of tau}. The tuning parameter for regularization in \cite{demiguel2009generalized} is selected via cross-validation with an out-of-sample risk targeting loss function. The target portfolio return is set to 0.05\% per day.

\subsection{Data}

We conduct our empirical studies based on daily excess return data of the S\&P 500 index component stocks collected from the CRSP(Center for Research in Security Prices). We delete those stocks that do not have complete historical data at the first decision point.  To avoid look-ahead bias due to missing data during the rolling window process, we simply drop the corresponding assets from the basket \citep{ao2019approaching,fan2022}. The risk-free data are downloaded from the Fama-French data library.

We consider the following three scenarios: (1) the largest
$ p = 100$ stocks (measured by market value) with sample size  $T = 125$  (half year); 
(2) the largest $p = 100$ stocks with sample size $T = 250$  (one year) historical data; (3) the largest $p = 200$ stocks with sample size $T = 250$ periods. The holding period $T_h$ is set to 5 (weekly position rebalance) and 21 days (monthly position rebalance). We consider a long-term investment
practice that the out-of-sample period is from 03/01/2000 to 31/12/2019.

\subsection{Results}

Table~\ref{Table empirical results} reports the empirical results of three considered scenarios. It is evident that the proposed strategy HD-DRO performs well in practical application. First, the HD-DRO achieves the lowest out-of-sample risk among all portfolios in all cases. These results are desirable for a "robust" portfolio and are consistent with our simulation results. More importantly, the HD-DRO strategy has a good return-risk balance and it is the best overall performer in the SR. The SRs of HD-DRO are close to 0.04 in all cases, whereas the second-rank strategy often has SR less than 0.035. Furthermore, the HD-DRO strategy also has the lowest maximum drawdown (close to 40\%) in most cases, which also implies the robustness of HD-DRO portfolios. For the main competitor DRO strategy, it is clear that its performance is close to the 1/N strategy with slight superiority in terms of SR and risk. It is reasonable since 1/N strategy is the special case of DRO strategy with $\delta = \infty$, whereas DRO itself takes the optimal value of $\delta$. The worst performance is often observed in MV or MV-L1 strategies from the SR view, with the latter generally having a much lower risk than the former. These results are expected, as the MV strategy is not suitable for high-dimensional set-ups due to the applied sample covariance matrix, and we choose the tuning parameter for MV-L1 by minimizing the out-of-sample risk. For MV-POET, although it often achieves the largest cumulative excess return, it also undertakes the highest risk, and thus its SR is still smaller than that of our proposal.
\begin{table}[!htb]
\centering
\tabcolsep 0.13in
\caption{Empirical results of considered three scenarios. We report the cumulative excess return (CR), the standard deviation of the portfolio excess return (Risk), the Sharpe ratio (SR) and the Maximum drawdown (MDD).}
\label{Table empirical results}
\begin{tabular}{@{}lccccccccc@{}}
\toprule
        & CR      & Risk     & SR      & MDD      & \multicolumn{1}{c}{} & CR     & Risk     & SR      & MDD      \\ \cmidrule{2-5} \cmidrule{7-10} 
        & \multicolumn{4}{c}{monthly rebalancing} &                      & \multicolumn{4}{c}{weekly rebalancing} \\ \cline{2-5} \cmidrule{7-10} 
        & \multicolumn{4}{c}{$p = 100, T = 125$}    &                      & \multicolumn{4}{c}{$p = 100, T = 125$}   \\ \midrule
HD-DRO  & 1.629   & 0.00810  & 0.0401  & 39.95\%  &                      & 1.591  & 0.00794  & 0.0398  & 44.41\%  \\
DRO     & 1.429   & 0.01221  & 0.0233  & 74.28\%  &                      & 1.386  & 0.01216  & 0.0227  & 78.09\%  \\
$1/N$     & 1.562   & 0.01375  & 0.0226  & 87.73\%  &                      & 1.576  & 0.01374  & 0.0228  & 87.61\%  \\
MV      & 1.142   & 0.01216  & 0.0187  & 77.55\%  &                      & 1.323  & 0.01199  & 0.0219  & 67.22\%  \\
MV-L1   & 0.769   & 0.00841  & 0.0182  & 57.66\%  &                      & 0.880  & 0.00824  & 0.0212  & 62.06\%  \\
MV-POET & 2.440   & 0.01329  & 0.0366  & 42.66\%  &                      & 2.091  & 0.01302  & 0.0319  & 40.21\%  \\ \midrule
        & \multicolumn{4}{c}{$p = 100, T = 250$}    &                      & \multicolumn{4}{c}{$p = 100, T = 250$}   \\
HD-DRO  & 1.671   & 0.00818  & 0.0407  & 33.59\%  &                      & 1.697  & 0.00799  & 0.0422  & 34.13\%  \\
DRO     & 1.302   & 0.01129  & 0.0230  & 69.52\%  &                      & 1.272  & 0.01118  & 0.0226  & 71.00\%  \\
$1/N$     & 1.562   & 0.01375  & 0.0226  & 87.73\%  &                      & 1.576  & 0.01374  & 0.0228  & 87.61\%  \\
MV      & 1.278   & 0.00909  & 0.0280  & 55.69\%  &                      & 1.211  & 0.00885  & 0.0272  & 56.24\%  \\
MV-L1   & 0.884   & 0.00833  & 0.0211  & 52.55\%  &                      & 0.974  & 0.00821  & 0.0236  & 51.99\%  \\
MV-POET & 1.482   & 0.01312  & 0.0225  & 64.40\%  &                      & 2.609  & 0.01596  & 0.0325  & 101.60\% \\ \midrule
        & \multicolumn{4}{c}{$p = 200, T = 250$}    &                      & \multicolumn{4}{c}{$p = 200, T = 250$}   \\
HD-DRO  & 1.389   & 0.00758  & 0.0365  & 42.49\%  &                      & 1.560  & 0.00740  & 0.0419  & 35.30\%  \\
DRO     & 1.559   & 0.01179  & 0.0263  & 73.73\%  &                      & 1.530  & 0.01173  & 0.0259  & 78.63\%  \\
$1/N$     & 1.758   & 0.01321  & 0.0265  & 87.23\%  &                      & 1.781  & 0.01322  & 0.0268  & 87.15\%  \\
MV      & 0.913   & 0.01050  & 0.0173  & 67.49\%  &                      & 1.247  & 0.01042  & 0.0238  & 75.30\%  \\
MV-L1   & 0.733   & 0.00774  & 0.0189  & 58.34\%  &                      & 0.793  & 0.00761  & 0.0207  & 58.81\%  \\
MV-POET & 2.286   & 0.01270  & 0.0359  & 64.57\%  &                      & 2.017  & 0.01265  & 0.0317  & 60.35\% \\
\bottomrule
\end{tabular}
\end{table}

\begin{figure}[!htb]
\centering	
\includegraphics[width=12cm, trim={0cm 0 0 0}]{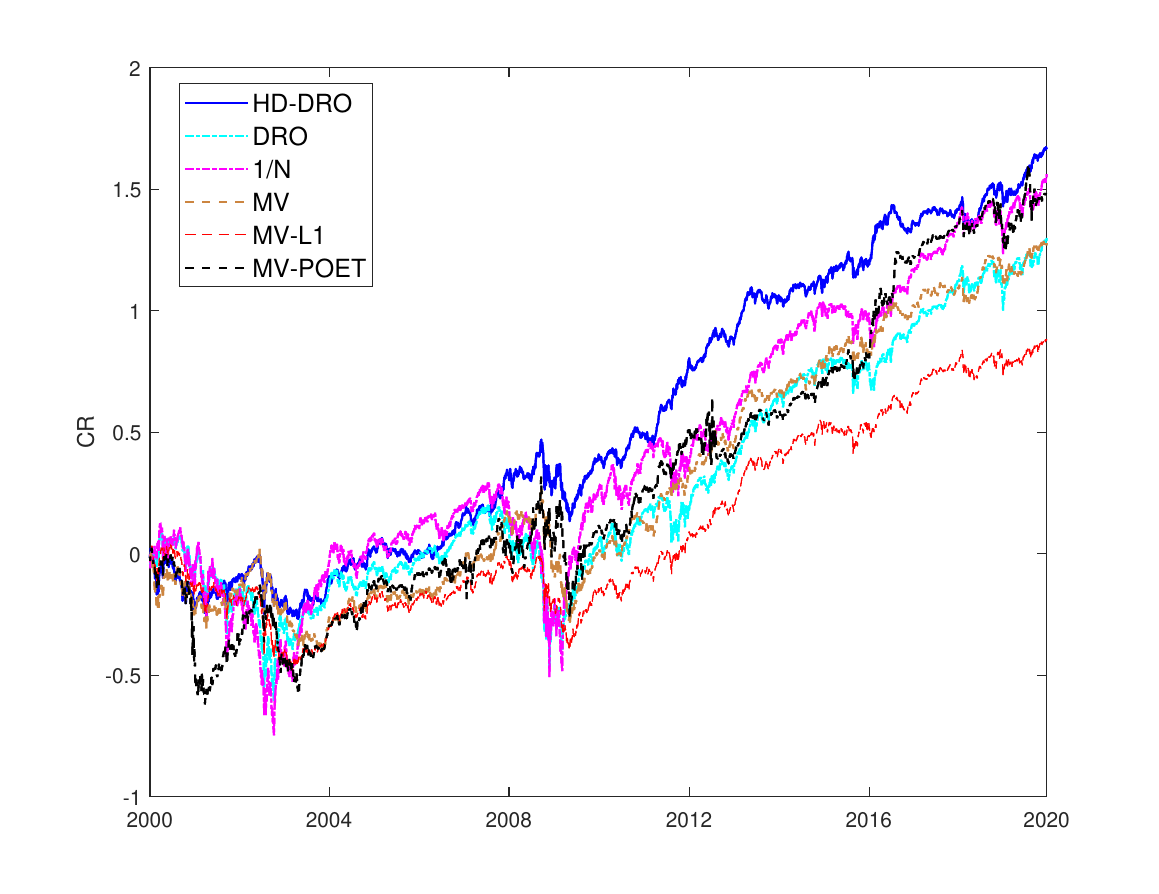}	
\caption{Time plot of the cumulative excess return of the considered strategies in the case of the largest 100 stocks with sample size $T=250$ and monthly rebalancing.}\label{Figure empirical CR density}
\end{figure}

Figure~\ref{Figure empirical CR density} plots the cumulative excess returns of all strategies in scenario (2) with monthly rebalancing. It is evident that during the 2008 financial crisis, except for HD-DRO, all other portfolios experienced a large decline in returns. It implies the robustness of HD-DRO.
Furthermore, the DRO and 1/N portfolios exhibit very similar performance in most periods, with their lines often running parallel to each other.

To demonstrate the superiority of the HD-DRO strategy, we evaluate the performance of all mentioned strategies when the stocks in the investment pool are randomly selected from S\&P 500 components. Specifically, the out-of-sample period is still from 03/01/2000 to 31/12/2019. We randomly draw 100 stocks from the S\&P 500 components to construct various portfolios. Then, we repeat the rolling window experiments on 100 such sets of 100 stocks. The holding period and sample size are fixed at 21 (monthly re-balance) and 250 (one year).

\begin{figure}[!htb]
\centering	
\subfloat[Empirical density of the standard deviation of portfolio excess return]
{\includegraphics[width=8.3cm]{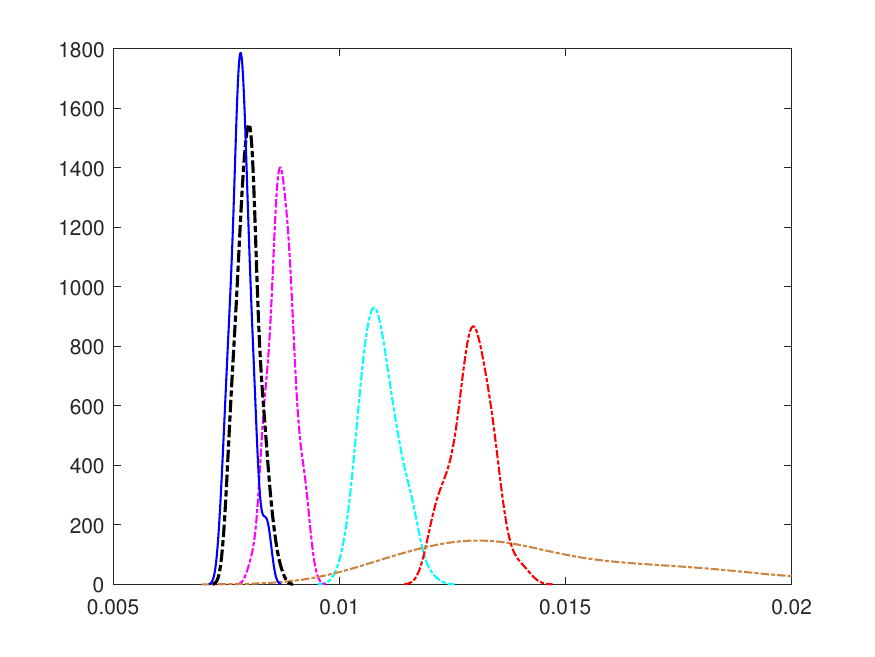}\label{Figure empirical risk density}}
\quad
\subfloat[Empirical density of  portfolio Sharpe ratio]
{\includegraphics[width=8.3cm]{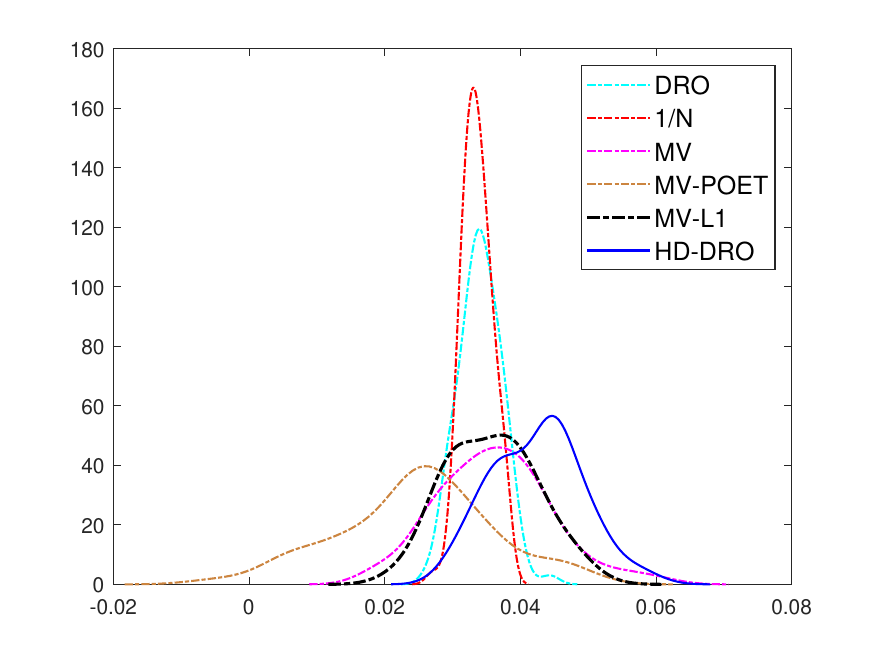}\label{Figure empirical SR density}}
\caption{The empirical density of portfolio standard deviation and Sharpe ratio over $100$ random experiments.}
\end{figure}

Figure~\ref{Figure empirical risk density} shows the empirical density of portfolio standard deviation (risk).  It is clear that the HD-DRO strategy achieves the lowest risk in most situations. Also we can observe that the maximum HD-DRO portfolio risk over 100 experiments is still much smaller than the risk of the DRO portfolio, which shows the significant advantages of HD-DRO over DRO strategy. Further, turn to Figure~\ref{Figure empirical SR density}, which plots the empirical density of the SR of various portfolios. It is evident that HD-DRO has  shifted  more to the right than other competing strategies, indicating the former outperforms the other strategies in terms of SR.  Additionally, the DRO and 1/N are almost equally concentrated. The SR of MV-POET does not show significant superiority when stocks are randomly selected as it does in the largest stock cases reported in Table~\ref{Table empirical results}.

\section{Conclusion}\label{sec:8}

In this paper, we study distributionally robust portfolio allocation based on  factor model in high-dimensional situations. The model uncertainty is measured by the Wasserstein distance. Specifically, we impose the uncertainty set, the region centred by empirical probability measure with radius $\delta$, on the common factors that drive the co-movement of assets return when considering the mean-variance portfolio allocation. The distributionally robust model is a min-max problem that is hard to solve. Thus, we reform it to a non-robust optimization problem with regularization via a dual argument. The problem involves two parameters $\delta$ and $\rho$, which measure the size of uncertainty set and the lowest acceptable portfolio return. Both parameters are crucial for the portfolio allocation. Then we discuss the data-driven procedures for choosing each of them. The optimal choices of $\delta$ and $\rho$ require knowledge of some unknown population quantities. Therefore, we provide further estimation procedures and demonstrate the corresponding asymptotic convergence. The Monte Carlo simulation shows that the estimated values of optimal $\delta$ and $\rho$ are close to the corresponding oracle version, and our distributionally robust portfolio enjoys low risk.  Finally, we conduct the empirical analysis based on S\&P 500 index components from 01/2000 to 12/2019. Compared to the benchmark strategies considered, particularly the distributionally robust portfolio of BCZ, the proposed portfolio demonstrates better performance in terms of SR and risk measures.

\section*{Acknowledgment}

The first author acknowledges research support from China Scholarship Council. This project was partially supported by a research grant from the Data Horizons Consilience Centre at Macquarie University.

\bibliographystyle{chicago}
\bibliography{reference}

\newpage
{\Large \bf 
		\begin{center}
			Supplementary Material to ``Uncertainty Learning for High-dimensional Mean-variance Portfolio''
		\end{center}
	}
	{
		\begin{center}
			{\sc Ruike Wu}$^{\dag}$, {\sc Yanrong  Yang}$^{\ddag}$, {\sc Han Lin Shang }$^{\sharp}$, {\sc and Huanjun Zhu}$^{\dag}$\\
			{Xiamen University$^{\dag}$}\\
			{The Australian National University $^{\ddag}$} \\ 
			{Macquarie University$^\sharp$} 
		\end{center}
	}

\appendix

\setcounter{page}{1}
\counterwithin{figure}{section}
\counterwithin{table}{section}

This supplementary material provides the detailed proofs of the main results in the paper.

\section{The main proof of the paper}

\subsection{Proof of Theorem 1}

Let's first recall the distributionally robust mean-variance problem, which is given by 
\begin{align}
\mathop{\min}_{w}  \left\{ \left[ \mathop{\max}_{\mathbb{P} \in \mathcal{U}_\delta(\mathbb{P}_T)}   w^\top B\text{Var}_{\mathbb{P}}({F}_t)B^\top w \right ] + w^\top \Sigma_{e}^* w \right\}\label{Appendix: our basic model}
\\
\text{s.t.} \quad w^\top \textbf{1}_p = 1, \; \min_{\mathbb{P}\in \mathcal{U}_\delta(\mathbb{P}_T)}\text{E}_{\mathbb{P}}(w^\top B {F}_t) \geq \rho. \notag
\end{align}

Since the number of common factors $K$ is finite, we can directly apply proposition~A.1. of \cite{blanchet2022distributionally} and derive that 
$$
\min_{\mathbb{P}\in \mathcal{U}_\delta(\mathbb{P}_T)}{\text{E}}_{\mathbb{P}}(w^\top B {F}_t) = \text{E}_{\mathbb{P}_T}(w^\top BF_t) - \sqrt{\delta}\|  B^\top w \|_q.
$$

Therefore, the feasible region of \eqref{Appendix: our basic model} can be expressed as 
\begin{align}
\mathcal{F}_{\delta,\rho}(T) =  \{ w\in \mathbb{R}^p: w^\top\textbf{1}_p = 1,  E_{\mathbb{P}_T}(w^\top BF_t) \geq \rho + \sqrt{\delta}\|  B^\top w \|_q \}.\label{appendix:feasible set}
\end{align}
Note that $\mathcal{F}_{\delta,\rho}(T)$ is convex. 

We now consider the maximization problem:
\begin{align}
\mathop{\max}_{\mathbb{P} \in \mathcal{U}_\delta(\mathbb{P}_T)} \left\{ w^\top B\text{Var}_{\mathbb{P}}({F}_t)B^\top w \right\}.
\label{appendix: inner max 1}
\end{align}
Let $\tilde{w} = B^\top w \in \mathbb{R}^K$, by fixing $\text{E}_{\mathbb{P}}(\tilde{w}^\top F_t) = \overline{\rho} \geq \rho$ in the maximization problem, we have the following equivalent formulation for \eqref{appendix: inner max 1}:
\begin{align}
\max_{\overline{\rho}\geq \rho}\left[\mathop{\max}_{\mathbb{P} \in \mathcal{U}_\delta(\mathbb{P}_T), \text{E}_{\mathbb{P}}(\tilde{w}^\top F_t) = \overline{\rho}}  \left\{ \tilde{w}^\top \text{E}_{\mathbb{P}}({F}_tF_t^\top)\tilde{w}\right\} - \overline{\rho}^2\right]. 
\label{appendix: inner max1 double}
\end{align}
By Proposition A.2 of \cite{blanchet2022distributionally}, the optimal value function of  
$$\mathop{\max}_{\mathbb{P} \in \mathcal{U}_\delta(\mathbb{P}_T), \text{E}_{\mathbb{P}}(\tilde{w}^\top F_t) = \overline{\rho}}  \left\{ \tilde{w}^\top \text{E}_{\mathbb{P}}({F}_tF_t^\top)\tilde{w}\right\}$$ 
is given by 
\begin{align}
\inf_{\kappa_1 \geq 0 ,\kappa_2}\left[\frac{1}{T}\sum_{t=1}^T \Phi(F_t) + \kappa_1 \delta + \kappa_2 \overline{\rho} \right],
\end{align}
where
$$
\Phi(F_t) := \sup_u \left[ (\Tilde{w}^\top u)^2 - \kappa_1 \| u-F_t \|_\iota^2 - \kappa_2\Tilde{w}^\top u \right].
$$
Furthermore, by applying Proposition A.3 of \cite{blanchet2022distributionally}, if $\left(\overline{\rho} - \tilde{w}^{\top} \text{E}_{\mathbb{P}}(F_t) \right)^2 - \delta \|\tilde{w} \|_\iota^2\leq 0$, the value of \eqref{appendix: inner max 1} is equal to 
\begin{align}
h(\overline{\rho},\tilde{w}) :=& \text{E}_{\mathbb{P}_T}\left[ (\tilde{w}^\top F_t)^2 \right] + 2\left(\overline{\rho} - \tilde{w}^\top \text{E}_{\mathbb{P}_T}(F_t) \right)\tilde{w}^\top \text{E}_{\mathbb{P}_T}(F_t) + \delta\| \tilde{w}\|_q^2 \notag \\&+ 2\sqrt{\delta \| \tilde{w}\|_2^2-\left(\overline{\rho} - \Tilde{w}^\top \text{E}_{\mathbb{P}_T}(F_t) \right)^2}\sqrt{\tilde{w}^\top \text{Var}_{\mathbb{P}_T}(F_t) \Tilde{w}} .
\end{align}
where $1/\iota + 1/q = 1$. Then, following the same proving procedure as those for Theorem~1 in \cite{blanchet2022distributionally}, the optimal value of~\eqref{appendix: inner max1 double} is given by 
\begin{equation}
\left( \sqrt{\Tilde{w}^{\top} \text{Var}_{\mathbb{P}_{T}}(F_t)\Tilde{w}_t} + \sqrt{\delta}\|\tilde{w}\|_q\right)^{2}.\label{appendix: factor part inner}
\end{equation}
As a result, we can replace \eqref{appendix: inner max 1} in \eqref{Appendix: our basic model} with \eqref{appendix: factor part inner}, and problem \eqref{Appendix: our basic model} can be reformed as 
\begin{align*}
\mathop{\min}_{w} \left(\sqrt{w^{\top} B \text{Var}_{\mathbb{P}_T}(F_t) B^\top w}  + \sqrt{\delta} \| B^\top w\|_q \right)^2 + w^\top \Sigma_{e}^* w \\
\text{s.t.} \ w^\top \bm{1}_p = 1, w^{\top} B \text{E}_{\mathbb{P}_T}(F_{t}) \geq \rho + \sqrt{\delta}||B^\top w||_{q}, \notag
\end{align*}

$\Box$

\subsection{Proof for Theorem 2}

Recall the logic of choosing $\delta$ in Subsection 3.1 of the main paper, the value of $\delta$ is given by the $1-\alpha_0$ quantile of the limit of $\mathcal{R}_{T}(\lambda_T,\mu_{f}^{*} ,\mathcal{S})$, where $\mu_f^{*} = \text{E}_{\mathbb{P}_T}(F_t )$, $\mathcal{S} = \text{E}_{\mathbb{P}_T}(F_tF_t^\top)$ and $\lambda_T = \text{E}_{\mathbb{P}_T}\left(2\phi_f^* \left( F_t F_t^\top -\mu_f^*\mu_f^{*\top} \right) \phi_f^*\right)$. Let
\begin{align}
R_T(\lambda,\mu, \Sigma) := \inf\Big\{ D_c(\mathbb{P},\mathbb{P}_T):\; & \text{E}_{\mathbb{P}}(F_t-\mu) = 0, \; \text{E}_{\mathbb{P}}(F_tF_t^\top-\Sigma) = 0, \notag\\
&\text{E}_{\mathbb{P}}\left(2\phi_f^{*\top}\left( F_t F_t^\top - \mu_f^* \mu_f^{*\top} \right) \phi_f^* - \lambda\right) = 0 \Big\},
\end{align}
where $D_c(\mathbb{P},\mathbb{P}_T)$ is the Wasserstein distance\footnote{For easier reading, let's recall the definition of the optimal transport costs and Wasserstein distances. Let $c:\mathbb{R}^p\times\mathbb{R}^p\to[0,\infty]$ be any lower semi-continuous function such that $c(u,u)=0$ for every $u \in\mathbb{R}^p$ . Given two probability distributions ${P}(\cdot)$ and ${Q}(\cdot)$ supported on $\mathbb{R}^p$, the optimal transport cost or discrepancy between P and Q, denoted by $D_{c}(P, Q)$, is defined as
\begin{align*}
{D}_{c}(\mathrm{P},\mathrm{Q})=\operatorname*{inf}\{\mathrm{E}_{\pi}[c(U,W)]:\pi\in\mathcal{P}(\mathbb{R}^{p}\times\mathbb{R}^{p}),\:\pi_{U}=\mathrm{P},\:\pi_{W}=\mathrm{Q}\}.
\end{align*}
Here, $P(\mathbb{R}^p\times\mathbb{R}^p)$ is the set of joint probability distributions $\pi$ of $(U,W)$ supported on $\mathbb{R}^p\times$ $\mathbb{R}^p$, and $\pi_U,\pi_W$ denote the marginal distributions of $U$ and $W$ , respectively, under the joint distribution $\pi$ . Intuitively, the quantity $c(u,\phi)$ can be interpreted as the cost of transporting unit mass from $u$ in $\mathbb{R}^p$ to another element $\phi$ in $\mathbb{R}^p$. Then the expectation $\mathbb{E}_{\pi}[c(U, W)]$ corresponds to the expected transport cost associated with the joint distribution $\pi$. In this paper, we set $c(u,\phi) = || u -\phi ||_\iota^2$ for simplicity.} 
 between $\mathbb{P}$ and $\mathbb{P}_T$.
Since we choose that $\lambda_T = 2\phi^{*\top}_f\left(\mathcal{S} -\mu_f^*\mu_f^{*\top} \right)\phi_f^*$, we have $\mathcal{R}_T(\lambda_T, \mu_f^*, \mathcal{S}) = {R}_T(\lambda_T, \mu_f^*, \mathcal{S})$. As a result, it is enough to derive the limit of ${R}_T(\lambda, \mu, \Sigma)$ as $T\rightarrow \infty$ for any symmetric covariance matrix $\Sigma$.
By using the definition of Wasserstein distance,  we can reform $R_T(\lambda,\mu, \Sigma)$ as 
\begin{align}
R_T(\lambda, \mu ,\Sigma) = \inf\Big\{&\text{E}_\pi\left(\| U-F_t\|_\iota^2 \right): \; \text{E}_{\pi} (U-\mu) = \textbf{0}_K, \; \text{E}_{\pi}(UU^{\top}-\Sigma) = \textbf{0}_{K\times K}, \;\pi_f = \mathbb{P}_T, \notag\\ 
&  \text{E}_\pi\left( 2\phi_f^{*\top} \left(UU^\top -\mu_f^* \mu_f^{*\top} \right) \phi_f^* -\lambda \right) = 0, \pi \in \mathcal{P}(\mathbb{R}^K\times \mathbb{R}^K)\Big\},\label{eq: RT defin trans}
\end{align}
where $\pi_f$ is the marginal distribution of $F_t$ and $\mathbb{P}_T$ is the empirical distribution formed from distinct samples $\{F_1,\ldots,F_T \}$, $\textbf{0}_d$ and $\textbf{0}_{d_1\times d_2}$ are d-dimensional zero column vector and $d_1\times d_2$ zero matrix, respectively. Now we are going to derive the dual representation of $R_T(\lambda, \mu, \Sigma)$. Before doing that, it is useful to introduce the concept of "the problem of moments" and the related proposition \ref{prop 1}. The concept of "the problem of moments" is mentioned by \cite{blanchet2019robust} in their Appendix B, and we provide a brief introduction here to make this section self-contained.

\textbf{The problem of moments.}
Let  $\Omega$  be a nonempty Borel measurable subset of  $\mathbb{R}^{p}$, which, in turn, is endowed with the Borel sigma algebra  $\mathcal{B}_{\Omega}$. Let  $X$  be a random vector taking values in the set  $\Omega$, and  $f=\left(f_{1}, \ldots, f_{k}\right): \Omega \rightarrow \mathbb{R}^{k}$  be a vector of moment functionals. Let  $\mathcal{P}_{\Omega}$  and  $\mathcal{M}_{\Omega}^{+} $ denote, respectively, the set of probability and non-negative measures, respectively on  $\left(\Omega, \mathcal{B}_{\Omega}\right)$  such that the Borel measurable functionals  $\phi, f_{1}, f_{2}, \ldots, f_{k}$, defined on  $\Omega$, are all integrable. Given a real vector  $q=\left(q_{1}, \ldots, q_{k}\right)$, the objective of the problem of moments is to find the worst-case bound,
\begin{equation}
v(q):=\sup \left\{\mathbb{E}_{\mu}[\phi(X)]:\; \mathbb{E}_{\mu}[f(X)]=q,\; \mu \in \mathcal{P}_{\Omega}\right\}.\label{eq: v(q) first}
\end{equation}

If we let $f_{0}=\mathbf{I}_{\Omega}$ where $\mathbf{I}$ is the indicator function, we have  $\mathbb{E}_{\mu}\left[f_{0}(X)\right]=1 $. Then, we define  $\tilde{f} := \left(f_{0}, f_{1}, \ldots, f_{k}\right), \tilde{q} :=\left(1, q_{1}, \ldots, q_{k}\right)$, and consider the following reformulation of the above problem:
\begin{equation}
v(q):=\sup \left\{\int \phi(x) d \mu(x): \int \tilde{f}(x) d \mu(x)=\tilde{q}, \mu \in \mathcal{M}_{\Omega}^{+}\right\}.\label{eq:dual representaion of v(q)}
\end{equation}
Based on the reformulation~\eqref{eq:dual representaion of v(q)}, under the assumption that a certain Slater's type of condition is satisfied, the following Proposition~\ref{prop 1} gives the equivalent dual representation for $v(q)$. 

\begin{prop}(Theorem~1 of \cite{keisii1962sharpness}).
\label{prop 1}
Let $Q_{\tilde{f}}=\left\{\int\tilde{f}(x)d\mu(x):\mu\in\mathcal{M}_{\Omega}^{+}\right\}.If\:\tilde{q}=(1,q_1,\ldots,q_k)$
is an interior point
of $\mathcal{Q}_{\tilde{f}}$, then
\begin{equation}
v(q) =\inf_{a_i\in R} \left\{\sum_{i = 0}^k a_iq_i: \sum_{i=0}^k a_i \Tilde{f}_i(x) \geq \phi(x), \ \forall x\in\Omega\right\}.\label{eq: moment problem}
\end{equation}
\end{prop}
The proof of Proposition~\ref{prop 1} can be referred to the Theorem~1 of \cite{keisii1962sharpness} and its application [I]. We observe that the proof relies solely on mathematical algebra, so the result also applies as $k$ approaches infinity. By Proposition~\ref{prop 1}, the $v(q)$ in expression~\eqref{eq:dual representaion of v(q)} can be reformulated as the form of ~\eqref{eq: moment problem}. 

Now, we turn back to our target equation~\eqref{eq: RT defin trans}.
Let $\Omega = \{ (u,F) \in \mathbb{R}^K \times \{F_1,\ldots,F_T \} : \|u-F \|_\iota^2 <\infty \}$, and
$$
f(u,F) = \left(\begin{array}{c}
    \textbf{I}_{\{F = F_1\}}(u,F)  \\
     \textbf{I}_{\{F = F_2\}}(u,F) 
     \\ \vdots
     \\
     \textbf{I}_{\{F = F_T\}}(u,F)
     \\
     u -\mu
     \\
     \text{vec}(u u^\top - \Sigma) 
    \\
    2\phi_f^{*\top}\left(u u^\top - \mu_{f}^{*} \mu_f^{*\top} \right) \phi_f^* -\lambda
\end{array} \right), \quad q= \left( \begin{array}{c}
     1/T  \\
     1/T \\ \vdots\\ 1/T
     \\
     \textbf{0}_K\\\textbf{0}_{K(K+1)/2} 
     \\
     0
\end{array}\right).
$$
where $\text{vec}(M)$ rearranges the elements of the upper triangle of the symmetric matrix $M$ into a column vector. Then, $\forall (u,F) \in \Omega$, we have 
\begin{align}
R_T(\lambda, \mu, \Sigma)=  - \sup \Big\{\text{E}_\pi \left( - \| U-F_t\|_\iota^2 \right): \; \text{E}_\pi(f(U,F_t)) = q, \; \pi \in \mathcal{P}_\Omega \Big\}.\label{eq: RT reform 2}
\end{align}
where $\text{E}_\pi(f(U,F_t)) = q$ corresponds to the constraints $\pi_f = \mathcal{P}_T$, $\text{E}_\pi(U-\mu) = \textbf{0}_K$, $\text{E}_\pi(UU^\top - \Sigma) = \textbf{0}_{K\times K}$ and $\text{E}_\pi(2\phi^{*\top}_f \left(UU^\top -\mu_f^* \mu_f^{*\top} \right) \phi^*_f - \lambda) = 0$. We note that  the constraints $\text{E}_\pi(\textbf{I}_{\{F = F_i\}}(u,F)) = 1/T$ for $i = 1,\ldots, T$ together specify that $P_\pi(\Omega) = 1$, the constraint that $\text{E}_\pi (\textbf{I}_\Omega(u,F)) = 1 $ is redundant.  
In addition, let $\phi(u,F)= -\| U-F\|_\iota^2$, Eq.(\ref{eq: RT reform 2}) takes the same form as~\eqref{eq: v(q) first}, hence by the reformulation~\eqref{eq:dual representaion of v(q)}, we have
\begin{equation}
R_T(\lambda, \mu,\Sigma) = -\sup \Big\{ \int \phi(u,y)d\mu(u,y): \quad \int f(u,y)d\mu(u,y) = q, \quad \mu\in \mathcal{M}_\Omega^+\Big\},\label{appendix: moment formula us}
\end{equation}
where $\mathcal{M}_\Omega^{+}$ is the non-negative measures corresponding to the set of probability $\mathcal{P}_\Omega$. Moreover, as $\{ 0\}$ lies in the interior of convex hull of the range $\{u -\mu : (u,F)\in \Omega \}$, $\{2\phi_f^{*\top}\left(uu^\top-\mu_f^* \mu_f^{*\top}\right) \phi_f^* -\lambda: (u,F) \in \Omega \}$ and  $\{ uu^\top - \Sigma: (u,F)\in\Omega\}$, observe that the set $\mathcal{Q}_f:= \{ \int f d\mu : \mu \in \mathcal{M}_\Omega^+ \}$ is $\mathbb{R}_+^T \times \mathbb{R}$. Then the Slater's condition $q \in int(\mathcal{Q}_f)$ is satisfied for the moment problem~\eqref{appendix: moment formula us}. As a consequence, we can  apply  Proposition \ref{prop 1} and obtain that:
\begin{small}
\begin{align}
R_T(\lambda,\mu, \Sigma) =&  -\inf_{a_i \in \mathbb{R}} \left\{ \frac{1}{T}\sum_{i=1}^T a_i:\sum_{i=1}^Ta_i\textbf{I}_{\{F = F_i\}}(u,F)  + 
a_{T+K+(K+1)K/2+1} \left(2\phi_f^{*\top} \left(uu^\top -\mu_f^* \mu_f^{*\top}\right) \phi_f^* -\lambda \right) \right. \notag\\ 
& \left.  + \sum_{i=T+1}^{i= T+K} a_i(u_{i-T}-\mu_{i-T}) + \sum_{i=T+K+1}^{T+K+K(K+1)/2}a_{i}[\text{vec}(uu^\top -\Sigma)]_{i-K-T} 
\geq -\| u -F \|_\iota^2, \forall (u,F) \in \Omega \right\} \notag\\  
=& -\inf_{a_i \in \mathbb{R}, \Lambda\in \mathbb{R}^{K \times K}} \left\{ \frac{1}{T}\sum_{i=1}^T a_i:\sum_{i=1}^Ta_i\textbf{I}_{\{F = F_i\}}(u,F) + a_{T+K+(K+1)K/2+1} \left(2\phi_f^{*\top} \left(uu^\top -\mu_f^* \mu_f^{*\top}\right) \phi_f^* -\lambda \right)\right. \notag\\ 
& \left. + \sum_{i=T+1}^{i= T+K} a_i(u_{i-T}-\mu_{i-T}) + \text{tr}\left(\Lambda(uu^\top -\Sigma) \right)  \geq -\| u -F \|_\iota^2, \ for \ all (u,F) \in \Omega \right\} \notag\\  
=& -\inf_{\substack{
a_i \in \mathbb{R}, \Lambda \in \mathbb{R}^{K \times K}, \notag \\
\chi \in \mathbb{R}, \zeta \in \mathbb{R}^K
}} \Big\{ \frac{1}{T}\sum_{i=1}^T a_i: a_i \geq \sup_{u:\| u -F_i \|_\iota^2<\infty}\big\{ - \| u -F_i \|_\iota^2  - \chi\left(2\phi_f^{*\top} \left(uu^\top -\mu_f^* \mu_f^{*\top}\right) \phi_f^* -\lambda \right) \\
& - \zeta^\top (u-\mu) -\text{tr}\left(\Lambda(uu^\top -\Sigma) \right) \big\} \Big\}
\end{align}
\end{small}
\hspace{-.1in} where the second equality holds because the symmetry of $uu^\top - \Sigma$ ensures that  we can always find\footnote{For example, let $K = 3$  and  $\Lambda_{ij}$ be the $i$\textsuperscript{th} row and $j$\textsuperscript{th} column element of matrix $\Lambda$, then for arbitrary $A$, we can always let $\Lambda_{11} = a_{T+3+1}, (\Lambda_{12}+\Lambda_{21}) = a_{T+3+2}, \Lambda_{22} = a_{T+3+3},(\Lambda_{13}+\Lambda_{31}) = a_{T+3+4},(\Lambda_{23}+\Lambda_{32}) = a_{T+3+5}, \Lambda_{33} = a_{T+3+6}$ to make $\text{tr}\left(\Lambda(uu^\top - \Sigma) \right) = A^\top vec(uu^\top - \Sigma)$.}  a $\Lambda$ such that $\text{tr}\left(\Lambda(uu^\top - \Sigma) \right) = A^\top \text{vec}(uu^{\top} - \Sigma)$ where $A = (a_{T+K+1},\ldots,a_{T+K+ K(K+1)/2})^\top$. For the third equality, it holds since $(u,F_i) \in \Omega$ and the inner inequality is held for all $ (u,F) \in \Omega$, and we denote $\chi = a_{T+K+K(K+1)/2+1}$ and $\zeta = \left(a_{T+1},\ldots, a_{T+K}\right)^\top$ for notational convenience
. Note that the inner supremum is not affected when $\| u -F_i \|_\iota^2$ tends to positive infinity.
Consequently, we have
\begin{align}
    R_T(\lambda, \mu, \Sigma) = \sup_{ \Lambda \in \mathbb{R}^{K \times K},\chi\in \mathbb{R},\zeta \in \mathbb{R}^K}\left\{\frac{1}{T}\sum_{i=1}^T \inf_{u \in \mathbb{R}^K} \left\{\| u -F_i \|_\iota^2 + \zeta^\top (u-\mu) \right. \right. \notag
    \\ \left.\left. + \chi\left(2\phi_f^{*\top} \left(uu^\top -\mu_f^* \mu_f^{*\top} \right)\phi_f^* -\lambda \right)  + \text{tr}\left(\Lambda (uu^\top - \Sigma) \right) \right\} \right \}.
    \label{eq: RT middle 1}
\end{align}
Note that $\Lambda$, $\zeta$ and $\chi$ are all free variables,  and replace symbol '$i$' by symbol '$t$', \eqref{eq: RT middle 1} can be  rewritten as
\begin{align}
R_T(\lambda, \mu, \Sigma) = \sup_{ \Lambda \in \mathbb{R}^{K \times K},\chi\in \mathbb{R}, \zeta \in \mathbb{R}^K}\left\{ - \frac{1}{T}\sum_{t=1}^T \sup_{u \in \mathbb{R}^K} \left\{ \zeta^\top h_1(u,\mu)  + \text{tr}\left(\Lambda h_0(u,\Sigma)  \right)  \right. \right. \notag
\\  \left.\left. + \chi \left(2\phi_f^{*\top} \left(uu^\top -\mu_f^* \mu_f^{*\top} \right) \phi_f^* -\lambda \right)
- \| u -F_t \|_\iota^2 \right\} \right \}.\label{eq: dual reform}
\end{align}
where $h_0(u,\Sigma) = uu^\top -\Sigma$ and $h_1(u,\mu) = u-\mu$.
This completes the derivation for dual representation of $R_T(\lambda,
\mu, \Sigma)$.

Note that 
\begin{align*}
\mathop{\sup}_{u\in \mathbb{R}^K} & \left\{ \zeta^\top h_1(u,\mu)  + \text{tr}(\Lambda h_0(u,\Sigma))  + \chi \left(2\phi_f^{*\top} \left(uu^\top -\mu_f^* \mu_f^{*\top} \right)\phi_f^* -\lambda \right)  - ||u - F_t ||_\iota^2 \right\} 
\\ &= \mathop{\sup}_{v\in \mathbb{R}^K} \left\{ \zeta^\top h_1(F_t +v,\mu) + \text{tr}(\Lambda h_0(F_t+v,\Sigma)) + \chi\left(2\phi_f^{*\top} \left( (F_t + v)(F_t + v)^\top -\mu_f^* \mu_f^{*\top} \right) \phi_f^* -\lambda \right)   - ||v||_\iota^2 \right\} 
\\ &
= \mathop{\sup}_{v\in \mathbb{R}^K} \left\{ \text{tr}\left(\Lambda\left[ h_0(F_t+v,\Sigma) - h_0(F_t,\Sigma) \right]\right) + \zeta^\top v + 2\chi \phi_f^{*\top}(2F_t + v)v^\top \phi_f^*  - ||v||_\iota^2 \right\} 
\\ & + \zeta^\top (F_t -\mu)+  {\text{tr}}(\Lambda h_0(F_t,\Sigma)) + \chi\left( 2 \phi_f^{*\top}\left( F_tF_t^\top -\mu_f^* \mu_f^{*\top} \right) \phi_f^* - \lambda\right),
\end{align*}
where we let $v = u-F_t$ for the first equality, and use the relationship $\text{tr}\left(\Lambda h_0(F_t+v,\Sigma)   \right) = \text{tr}\left(\Lambda h_0(F_t+v,\Sigma) - \Lambda h_0(F_t,\Sigma)   \right) + \text{tr}\left(\Lambda h_0(F_t,\Sigma)   \right)$ for the second equality.  Furthermore, we can write 
\begin{align}
\text{tr}\left(\Lambda\left[ h_0(F_t+v,\Sigma) - h_0(F_t,\Sigma) \right]\right)  = \int_0^1 \frac{d}{dy}\text{tr}(\Lambda h_0(F_t+yv))dy. \label{eq: Tr decom 1}
\end{align}
Notice that 
\begin{align}
\frac{d}{dy}\text{tr}\left(\Lambda h_0(F_t+yv)\right) &= 
\frac{d}{dy} \text{tr}\left(\Lambda\left( (F_t+yv)(F_t+yv)^\top - \Sigma \right) \right) \notag
\\& = 2\text{tr}(\Lambda F_t v^\top) + 2yv^\top \Lambda v.    \label{eq: Tr decom 2}
\end{align}
Then combine (\ref{eq: Tr decom 1}) and (\ref{eq: Tr decom 2}), we obtain that
\begin{align}
\text{tr}\left(\Lambda\left[ h_0(F_t+v,\Sigma) - h_0(F_t,\Sigma) \right]\right)  = \int_0^1 \left( 2\text{tr}(\Lambda F_t v^\top) + 2yv^\top \Lambda v \right) dy = 2\text{tr}(\Lambda F_t v^\top) +  v^\top \Lambda v. \label{appendix: Tr - Tr}   
\end{align}
Further note that $\text{E}_{\mathbb{P}_T}(\text{tr}(\Lambda h_0(F_t,\mathcal{S}))) = 0$, $\text{E}_{\mathbb{P}_T}\left(\chi(2 \phi_f^{*\top}\left( F_tF_t^\top -\mu_f^*\mu_f^{*\top} \right) \phi_f^* - \lambda_T) \right) = 0$, and 
\begin{align*}
&4\chi \phi_f^{*\top}F_t v^\top \phi^{*}_{f} = \text{tr}\left( 4\chi \phi^*_f \phi_f^{*\top} F_t v^\top \right).
\end{align*}
Then we have
\begin{align*}
{R}_T(\lambda_T, \mu_f^*, \mathcal{S}) = &   \mathop{\sup}_{\zeta \in \mathbb{R}^K}\left\{\mathop{\sup}_{\Lambda \in \mathbb{R}^{K\times K}, \chi \in \mathbb{R}}  \left\{ -\frac{1}{T}\sum_{t=1}^T\left[ \mathop{\sup}_{\overline{v}}\left( \text{tr}\left( (2\Lambda  +  4\chi \phi^*_f \phi_f^{*\top}) F_t v^\top\right)  + \zeta^\top v \right.\right. \right. \right.
\\ & \left. \left.\left. \left.+ 2\chi \phi_f^{*\top}vv^\top \phi_f^* + v^\top \Lambda v
- ||v||_\iota^2 \right)  \right]   \right\} - \frac{1}{T}\sum_{t=1}^T\zeta^\top(F_t -\mu_f^*)\right\}.
\end{align*}

Now let $\overline{v} = {v}\sqrt{T}$, $\overline{\chi} = \chi\sqrt{T}$ and $\overline{\Lambda} = \Lambda\sqrt{T}$, and multiply $T$ on both sides, we obtain that
\begin{align}
T{R}_T(\lambda_T, \mu_f^*, \mathcal{S}) 
&=  \mathop{\sup}_{\overline{\zeta} \in \mathbb{R}^K}\left\{\mathop{\sup}_{\overline{\Lambda} \in \mathbb{R}^{K\times K}, \overline{\chi} \in \mathbb{R}} \left\{-\text{E}_{\mathbb{P}_T}\left[ \mathop{\sup}_{\overline{v}}\left( \text{tr}\left((2\overline{\Lambda} + 4\chi \phi_f^{*}\phi_f^{*\top} )F_t \overline{v}^\top\right) + \overline{\zeta}^\top\overline{v} 
\right. \right. \right. \right. \notag
\\ & \left. \left. \left. \left.
+ \frac{2\overline{\chi}\phi_f^{*\top}\overline{v}\overline{v}^\top \phi_f^*}{\sqrt{T}}+ \frac{\overline{v}^\top \overline{\Lambda} \overline{v}}{\sqrt{T}}  - ||\overline{v}||_2^2 \right)  \right]  \right\} - \frac{1}{\sqrt{T}}\sum_{t=1}^T\overline{\zeta}^\top(F_t -\mu_f^*)\right\} \nonumber
    \\ & \equiv \sup_{\overline{\zeta}\in \mathbb{R}^{K}}\left\{\sup_{\overline{\Lambda}\in \mathbb{R}^{K\times K}, \overline{\chi} \in \mathbb{R}} \left\{- \text{E}_{\mathbb{P_T}}\left(A_1(\overline{\Lambda},\overline{\chi},\overline{\zeta})\right)\right\} - \frac{1}{\sqrt{T}}\sum_{t=1}^T\overline{\zeta}^\top(F_t -\mu_f^*)\right\}.
    \label{eq: fianl pTR}
\end{align}
First consider the inner problem $A_1(\overline{\Lambda},\overline{\chi},\overline{\zeta})$. By using the same argument in \citet[][p.6408]{blanchet2022distributionally} based on Assumption 2, and note that {$\| \phi_f^*\| < \overline{C}$ for some positive constant $\overline{C}$}, we have that 
the terms  ${\overline{v}^\top \overline{\Lambda} \overline{v}}/{\sqrt{T}}$ and ${2\overline{\chi}\phi_f^{*\top}\overline{v}\overline{v}^\top \phi_f^*}/{\sqrt{T}}$ 
are asymptotically negligible, thus the optimal value of $A_1(\overline{\Lambda})$ is given by
\begin{align}
\mathop{\sup}_{\overline{v}}\left( -||\overline{v}||_\iota^2  + \text{tr}\left((2\overline{\Lambda} + 4\chi \phi_f^* \phi_f^{*\top}  )F_t \overline{v}^\top\right) + \overline{\zeta}^\top \overline{v} \right) &= \mathop{\sup}_{\overline{v}} \left\{ 2\left\vert\left\vert\left(\overline{\Lambda} + \overline{\chi} \phi_f^{*}\phi_f^{*\top} \right)F_t +\overline{\zeta}\right\vert\right\vert_q \left\vert\left\vert \overline{v}\right\vert\right\vert_\iota   - ||\overline{v}||_\iota^2 \right\} \nonumber
\\&= \left\| \left( \overline{\Lambda} + \overline{\chi} \phi_f^* \phi_f^{*\top}\right)F_t+\overline{\zeta}\right\|_q^2,
\label{eq: innermost opt}
\end{align}
where the first equality holds by H$\ddot{\mathrm{o}}$lder's inequality and $1/q + 1/\iota = 1$. 
As a result, we obtain:
\begin{align}
T{R}_T(\lambda_T, \mu_f^*, \mathcal{S}) = \mathop{\sup}_{\overline{\zeta}\in \mathbb{R}^K}\left\{ - \frac{1}{\sqrt{T}}\sum_{t=1}^T\overline{\zeta}^\top(F_t -\mu_f^*) -\mathop{\inf}_{\overline{\Lambda}\in \mathbb{R}^{K \times K},\chi\in\mathbb{R} } \left\{\text{E}_{\mathbb{P}_T}\left( \left\|\left(\overline{\Lambda}+ \overline{\chi} \phi_f^* \phi_f^{*\top}\right)F_t + \overline{\zeta}\right\|_q^2 \right)\right\} \right\}
    .\label{eq: pTR part3 reform}
\end{align}

Since the feasible regions for $\overline{\Lambda}$ and $\overline{\chi}$ are  $\mathbb{R}^{K\times K}$ and $\mathbb{R}$ respectively, then write $\Gamma = \overline{\Lambda} + \overline{\chi}\phi_f^*\phi_f^{*\top}$, 
the value of $\mathop{\inf}_{\overline{\Lambda}\in \mathbb{R}^{K \times K},\chi\in\mathbb{R} } \left\{\text{E}_{\mathbb{P}_T}\left( \left\|\left(\overline{\Lambda}+ \overline{\chi} \phi_f^* \phi_f^{*\top}\right)F_t + \overline{\zeta}\right\|_q^2 \right)\right\}$ is given by 
$$
\mathop{\inf}_{\Gamma\in \mathbb{R}^{K \times K}} \left\{ E_{\mathbb{P}_T}\left( \left\|\Gamma F_t + \overline{\zeta}\right\|_q^2 \right)\right\}.
$$
Based on the central limit theorem of Assumption 1 that $\frac{1}{\sqrt{T}}\sum_{t=1}^T (F_t - \mu_f^*) \overset{d}{\rightarrow} Z_0:= N(\textbf{0}_K, V_g)$ where $\mu_{f}^{*}= \text{E}_{\mathbb{P}^*}(F_t)$ and $V_g = \lim_{T\rightarrow\infty}\frac{1}{T}\sum_{t=1}^T\sum_{t^\prime = 1}^T\left( F_t -\mu_f^* \right)\left( F_{t^\prime} -\mu_f^* \right)^\top$. We have that 
\begin{align}
TR_T(\lambda_T,\mu_f^*,\mathcal{S}) \overset{d}{\rightarrow} L_0 := \mathop{\sup}_{\overline{\zeta}\in \mathbb{R}^K}\left\{ \overline{\zeta}^\top Z_0 - \mathop{\inf}_{\Gamma\in \mathbb{R}^{K \times K}} \left\{ \text{E}_{\mathbb{P}^*}\left( \left\|\Gamma F_t + \overline{\zeta}\right\|_{q}^{2} \right)\right\} \right\}. 
\end{align}

If $q = 2$, we have
$$
\text{E}_{\mathbb{P}_T}\left(\left\|\Gamma F_t+\overline{\zeta}\right\|_2^2\right) = \sum_{i=1}^{K} \text{E}_{\mathbb{P}_T}\left(\Gamma_iF_t + \overline{\zeta}_i\right)^2.
$$
Taking derivative with respect to each row of $\Gamma$, we obtain 
\begin{align}
\text{E}_{\mathbb{P}_T}(F_t F_t^\top) \Gamma_i^\top = - \overline{\zeta}_i \text{E}_{\mathbb{P}_T}(F_t). \label{appendix: foc second part}
\end{align}
 Plugging ~\eqref{appendix: foc second part} into $\text{E}_{\mathbb{P}_T}\left(\Gamma_iF_t + \overline{\zeta}_i\right)^2$ further gives that 
\begin{align}
\text{E}_{\mathbb{P}_T}\left(\Gamma_iF_t + \overline{\zeta}_i\right)^2  &= \Gamma_{i} \text{E}_{\mathbb{P}_T}(F_t F_t^\top) \Gamma_{i}^{\top} + 2 \overline{\zeta}_i \Gamma_i \text{E}_{\mathbb{P}_T}(F_t) + \overline{\zeta}_i^2  \nonumber \\ &= \overline{\zeta}_i^2\left(1-\text{E}_{\mathbb{P}_T}(F_t)^{\top} \left(\text{E}_{\mathbb{P}_T}(F_{t} F_{t}^{\top})\right)^{-1}\text{E}_{\mathbb{P}_T}(F_t)\right). 
\label{eq: foc iota 2 simply}  
\end{align}
Note that as $T\rightarrow \infty$, $\mathcal{S} = \text{E}_{\mathbb{P}_T}(F_tF_t^\top)$ is invertible asymptotically by Assumption 3. As a result, we derive that 
\begin{align}
\mathop{\inf}_{\Gamma} \left\{  \text{E}_{\mathbb{P}_T}\left(\left\|\Gamma F_t + \overline{\zeta}\right\|_2^2\right)\right\}  =   \ \left\vert\left\vert \overline{\zeta}  \right\vert \right\vert_2^2\left(1-\mu_T^\top \mathcal{S}^{-1}\mu_T\right), 
\label{eq: pTR second part final}
\end{align}
where $\mu_T = \text{E}_{\mathbb{P}_T}(F_t)$.
{Note that  $\mu_T^\top \mathcal{S}^{-1} \mu_T\overset{p}{\rightarrow} \mu_f^{*\top}\tilde{\Sigma}_f^{*-1}\mu_f^{*}$ when $K$ is fixed \citep[e.g. see][]{kan2007optimal}, where $\tilde{\Sigma}_f^* = \text{E}_{\mathbb{P}^*}(F_tF_t^\top)$}. Then, we have
\begin{align}
TR_T(\lambda_T,\mu_f^*,\mathcal{S}) \overset{d}{\rightarrow} L_0 := \mathop{\sup}_{\overline{\zeta}\in \mathbb{R}^K}\left\{ \overline{\zeta}^\top Z_0 -  \left\vert\left\vert \overline{\zeta}  \right\vert \right\vert_2^2\left(1-\mu_f^{*\top}\tilde{\Sigma}_f^{*-1}\mu_f^*\right) \right\}. 
\end{align}

Furthermore, by Assumption 3, the true covariance matrix of common factors $\Sigma_f^*$  is positive definite and thus it follows that 
\begin{align*}
\mu_f^{*\top}\tilde{\Sigma}_f^{*-1}\mu_f^* = \mu_f^{*\top}{\Sigma}_f^{*-1}\mu_f^* - \frac{\left(\mu_f^{*\top}{\Sigma}_f^{*-1}\mu_f^*\right)^2}{1 + \mu_f^{*\top}{\Sigma}_f^{*-1}\mu_f^*} = \frac{\mu_f^{*\top}{\Sigma}_f^{*-1}\mu_f^*}{1 + \mu_f^{*\top}{\Sigma}_f^{*-1}\mu_f^*} < 1,
\end{align*}
by using the Sherman-Morrison formula. As a result, 
\begin{align}
TR_T(\lambda_T,\mu_f^*,\mathcal{S}) \overset{d}{\rightarrow} L_0 = \frac{\| Z_0\|_2^2}{4\left( 1- \mu_f^{*\top}\tilde{\Sigma}_f^{*-1}\mu_f^*\right)}. 
\end{align}
$\Box$

\subsection{Basic Lemmas}

\begin{lemma}
(Theorems~3.1 and~3.2 of \cite{fan2013large}) Suppose $log p = o(T^{y/6})$, $y^{-1}= 3y_1^{-1}+1.5y_2^{-1}+y_3^{-3}+1$, $T = o(p^2)$, $\omega_T = \sqrt{1/p}+\sqrt{logp/T}$, and Assumptions 4 - 7 hold. Then, for a sufficiently large constant $C>0$ in the threshold $\tau_{ij}$, we have
$$
\| \widehat{\Sigma}_{e,sp} - \Sigma_e^*\| = O_p(\omega_T^{1-\varsigma} m_\varsigma) = o_p(1), \ \| \widehat{\Sigma}_{r} - \Sigma_r^*\|_{\max} = O_p(\omega_T) = o_p(1),
$$
where $m_\varsigma$ and $\omega_T$ are defined in Assumption 5. \label{lemma fan2013 theorem 3.1 .32}
\end{lemma}

\begin{lemma}\label{lemma: Theorem 3.3 of Poet}
(Theorem~3.3 of \cite{fan2013large}). Under the conditions of Lemma \ref{lemma fan2013 theorem 3.1 .32}. Let $\mathcal{V}$ denote the $K \times K$ diagonal matrix of the first $K$ largest eigenvalues of the sample covariance of $r_t$, $\varpi_T = T^{-1/2} + {T^{1/4}}{p^{-1/2}} $, and define $H = \frac{1}{T}\mathcal{V}^{-1}\widehat{F}^\top F B^\top B$. Then we have 
\begin{equation*}
\max_{i\leq p }\left\| \widehat{b}_i - Hb_i\right\| = O_p(\omega_T)=o_p(1), \quad \max_{t\leq T}\left\| \widehat{F}_t - HF_t\right\| = O_p\left(\varpi_T \right) = o_p(1).
\end{equation*}
\end{lemma}

\begin{lemma}\label{lemma: Lemma C.10 of Poet}
(Lemma C.10 of \cite{fan2013large}). Under the conditions of Lemma \ref{lemma fan2013 theorem 3.1 .32}, $\|H^\top H - I_K\|_F = O_p\left(p^{-1/2} + T^{-1/2} \right)$ and $\|H\| = O_p(1)$.    
\end{lemma}

\subsection{Proof of Theorem 3}

The convergence is a direct result by 
\begin{align}
\left| \widehat{\mu}_f^\top \widehat{\mu}_{f} - \mu_{f}^{*\top}\mu_{f}^{*} \right| &= \left|(\widehat{\mu}_f - H\mu_f^{*})^\top \widehat{\mu}_{f} +   \mu_f^{*\top}H^\top (\widehat{\mu}_f - H\mu_f^{*}) + \mu_f^{*\top}\left(H^\top H - I_K\right)\mu_{f}^{*}\right| \notag
\\& \leq
\left\|\widehat{\mu}_{f} - H\mu_{f}^{*}\right\|_{\max}\| \widehat{\mu}_f\|_{\max} +   \|\mu_f^{*\top}\|_{\max}\| H\| \|\widehat{\mu}_f - H\mu_{f}^{*}\|_{\max} + \|\mu_f^{*}\|_{\max}^{2} \|H^{\top} H-I_{K}\| \notag
\\ &
=O_p(T^{-1/2} + T^{1/4}p^{-1/2}) = o_p(1),
\end{align}
where $\|H\| = O_p(1)$ and $\| H^\top H- I_K\| = O_p(p^{-1/2}+T^{-1/2})$ by Lemma~\ref{lemma: Lemma C.10 of Poet},  $\|\mu_f^*\|_{\max}<C$ by Assumption 7(a), and
\begin{align}
\|\widehat{\mu}_f - H\mu_f^*\| &= \left\|\frac{1}{T}\sum_{t=1}^T\widehat{F}_t - \frac{1}{T}\sum_{t=1}^TH{F}_t + \frac{1}{T}\sum_{t=1}^TH{F}_t - H\mu_f^* \right\|\notag
\\ & \leq \max_t \| \widehat{F}_t - HF_t\| + H\frac{1}{T}\sum_{t=1}^T(F_t - \mu_f^*) \notag 
\\& = O_p(T^{-1/2} + T^{1/4}p^{-1/2}), \label{eq: appendix rate hatmu -mu}
\end{align}
where we apply Lemma~\ref{lemma: Theorem 3.3 of Poet} and the central limit theorem in Assumption 2 to determine the order of last equality.

\subsection{Proof of Theorem 4}

First, under the identification Assumption 8, the asymptotic results in Lemma \ref{lemma: Theorem 3.3 of Poet} can be simplified as follows:
\begin{equation}
\max_{i\leq p}\| \widehat{b}_i - b_i\| = O_p(\omega_T)=o_p(1), \quad \max_{t\leq T}\| \widehat{F}_t - F_t\| = O_p\left(\varpi_T\right) =o_p(1),\label{Lemma: asy for factor model}
\end{equation}
since $H = I_{K} + O_p(\Delta)$ and $\Delta = o(\sqrt{{\log p}/{T}}+{T^{1/4}}/{p^{1/2}})$. And, $\widehat{\mu}_f - \mu_f^* = o_p(1)$ by \eqref{eq: appendix rate hatmu -mu}.

We turn to the convergence of $\widehat{V}_{g, T}$.
Define $\tilde{V}_{g,T} = \tilde{\mathcal{C}}_T(0) + 2\sum_{j=1}^{T-1}k\left(\frac{j}{q_T}\right)\tilde{\mathcal{C}}_T(j)$ where $\tilde{\mathcal{C}}_T(j) = 
  \frac{1}{T}\sum_{t=j+1}^T\left(\widehat{F}_t -{\mu}_f^* \right)\left(\widehat{F}_{t-j} -{\mu}_f^* \right)^\top $ for $j\geq 0$. Due to the fact that  $\widehat{\mu}_f$ consistently converges to $\mu_f^*$, it is sufficient to prove the consistency of $\tilde{V}_{g, T}$ with respect to $V_{g}$. To see this, we make the following decomposition :
\begin{align}
\tilde{\mathcal{C}}_T(j) & =  \frac{1}{T}\sum_{t=j+1}^T\left(\widehat{F}_t - F_t + F_t -{\mu}_f^* \right)\left(\widehat{F}_{t-j} - F_{t-j} + F_{t-j} -{\mu}_f^* \right)^\top \notag\\
& = 
\frac{1}{T}\sum_{t=j+1}^T\left( F_t -{\mu}_f^* \right)\left( F_{t-j} -{\mu}_f^* \right)^\top 
+ \frac{1}{T}\sum_{t=j+1}^T\left(\widehat{F}_t - F_t  \right)\left(\widehat{F}_{t-j} - F_{t-j} \right)^\top \notag
\\ &
+ \frac{1}{T}\sum_{t=j+1}^T\left( F_t -{\mu}_f^* \right)\left(\widehat{F}_{t-j} - F_{t-j}  \right)^\top + \frac{1}{T}\sum_{t=j+1}^T\left(\widehat{F}_t - F_t \right)\left(F_{t-j} -{\mu}_f^* \right)^\top \notag
\\ & 
\equiv \tilde{\mathcal{C}}_{T,j1} + \tilde{\mathcal{C}}_{T,j2} + \tilde{\mathcal{C}}_{T,j3} + \tilde{\mathcal{C}}_{T,j4}.
\label{eq: Appendix Vg decompose}
\end{align}
Thus, $\tilde{V}_{g,T} = \tilde{V}_{g,T1}+\tilde{V}_{g,T2}+\tilde{V}_{g,T3}+\tilde{V}_{g,T4}$ where $\tilde{V}_{g,Ti} := \tilde{\mathcal{C}}_{T,0i} + 2\sum_{j=1}^{T-1}k(\frac{j}{q_T})\tilde{\mathcal{C}}_{T,ji}$ for $i=1,2,3,4$. First,
under Assumptions 1, 5, 6 and 9, it is shown that $\tilde{V}_{g,T1}\overset{p}{\rightarrow} V_g$ \citep[Theorem 2.1]{de2000consistency}. Then, for $\tilde{V}_{g,T2}$, we have  for all $ 0 \leq j \leq q_T$,
\begin{align*}
\left\|\tilde{\mathcal{C}}_{T,j2} \right\| =  \left\| \frac{1}{T}\sum_{t=j+1}^T\left(\widehat{F}_t - F_t  \right)\left(\widehat{F}_{t-j} - F_{t-j} \right)^\top \right\| = O_p\left((\max_t\|\widehat{F}_t - F_t \|)^2\right) = O_p\left(T^{-1} + \sqrt{T}p^{-1} \right)
\end{align*}
By triangular inequality, it further follows that $\|\tilde{V}_{g,T2}\| \leq \left(1 + 2\sum_{j=1}^{T-1}|k(j/q_T)| \right)\max_{0\leq j\leq q_T}\|\tilde{C}_{T,j2} \|  = O_p(q_TT^{-1}+q_T\sqrt{T}p^{-1}) = o_p(1)$ since $\lim sup_{T\rightarrow\infty} q_T^{-1}\sum_{j=1}^{T-1}|k(j/q_T)|<\infty$ \citep[Lemma 1]{jansson2002consistent} and $q_TT^{1/4}p^{-1/2} \rightarrow 0$. Similarly, for $\tilde{V}_{g,T3}$,  
we have 
\begin{align*}
\left\| \frac{1}{T}\sum_{t=j+1}^T\left({F}_t - \mu_f^* \right)\left(\widehat{F}_{t-j} - F_{t-j} \right)^\top \right\| \leq \left(\max_t\|\widehat{F}_t - F_t \|\right)\frac{1}{T}\sum_{t=j+1}^T\|F_t -\mu_{f}^{*}\| = O_p\left(T^{-1/2} + T^{1/4}p^{-1/2} \right),
\end{align*}
for all $j\leq q_T$,
and thus $\|\tilde{V}_{g,T3} \| = O_p\left(q_TT^{-1/2} + q_T T^{1/4}p^{-1/2} \right) = o_p(1)$. Similar to $\tilde{V}_{g,T3}$,  $\|\tilde{V}_{g,T4}\| = o_p(1)$.  As a result, we complete the proof of $\tilde{V}_{g,T} \overset{p}{\rightarrow} V_g$.

The rest work is to prove that $\widehat{V}_{g,T}$ converges to $\tilde{V}_{g,T}$. Taking decomposition as follow:
\begin{align*}
\widehat{\mathcal{C}}_T(j) &:= \frac{1}{T}\sum_{t=j+1}^T\left(\widehat{F}_t - {\mu}_f^* +\mu_f^* - \widehat{\mu}_f \right)\left(\widehat{F}_{t-j} -{\mu}_f^* +\mu_f^* - \widehat{\mu}_f\right)^{\top} \\ 
&= \tilde{\mathcal{C}}_T(j) + \frac{1}{T}\sum_{t=j+1}^T\left(\mu_f^* - \widehat{\mu}_f \right)\left(\widehat{F}_{t-j} -{\mu}_f^* \right)^\top \\ 
&+\frac{1}{T}\sum_{t=j+1}^T\left(\widehat{F}_t - {\mu}_f^* \ \right)\left(\mu_f^* - \widehat{\mu}_f\right)^\top + \frac{1}{T}\sum_{t=j+1}^T\left(\mu_f^* - \widehat{\mu}_f \right)\left( \mu_f^* - \widehat{\mu}_f\right)^\top
\\& \equiv \tilde{\mathcal{C}}_T(j) + \widehat{\mathcal{C}}_{T,2}(j) + \widehat{\mathcal{C}}_{T,3}(j) + \widehat{\mathcal{C}}_{T,4}(j)
,
\end{align*}
and $\widehat{V}_{g,T} = \tilde{V}_{g,T} + \widehat{V}_{g,T2}+\widehat{V}_{g,T3}+\widehat{V}_{g,T4}$ where $\widehat{V}_{g,Ti} := \widehat{\mathcal{C}}_{T,0i} + 2\sum_{j=1}^{T-1}k(\frac{j}{q_T})\widehat{\mathcal{C}}_{T,ji}$ for $i=2,3,4$. Note that $\|\widehat{\mu}_f -\mu_f^*\|_{\max} = O_p(\varpi_T)$.
As a result, following the similar steps for $\tilde{V}_{g,Ti},i=2,3,4$, we can conclude that $\widehat{V}_{g,T} = \tilde{V}_{g,T} +o_p(1) \overset{p}{\rightarrow} V_{g}$.  $\Box$

\subsection{Proof of Theorem 5}

To show consistency of $\widehat{w}_{mv}^\top \widehat{B} \widehat{V}_{g,T} \widehat{B}^\top \widehat{w}_{mv}$ with respect to ${w}_{mv}^{*\top} {B} {V}_{g} {B}^\top {w}_{mv}^*$, let us first consider 
\begin{align*}
\left|\widehat{w}_{mv}^\top \widehat{B} {V}_{g} \widehat{B}^\top \widehat{w}_{mv} - {w}_{mv}^{*\top} {B} {V}_{g} {B}^\top {w}_{mv}^*\right| & \leq  \left|\widehat{w}_{mv}^\top \widehat{B} {V}_{g} \widehat{B}^\top \widehat{w}_{mv} -   {w}_{mv}^{*\top} \widehat{B} {V}_{g} \widehat{B}^\top \widehat{w}_{mv}\right|
\\ & +
\left|{w}_{mv}^{*\top} \widehat{B} {V}_{g} \widehat{B}^\top \widehat{w}_{mv} -  {w}_{mv}^{*\top} {B} {V}_{g} {B}^\top \widehat{w}_{mv}\right|
\\& + 
\left|{w}_{mv}^{*\top} {B} {V}_{g} {B}^\top \widehat{w}_{mv} - {w}_{mv}^{*\top} {B} {V}_{g} {B}^\top {w}_{mv}^* \right|
\\& \leq \|\widehat{w}_{mv} - w_{mv}^* \|_{\max} \| \widehat{B}V_g\widehat{B}^\top\|_{max} \|\widehat{w}_{mv}\|_{\max}
\\ & + \|{w}_{mv}^*\|_{\max}\| \widehat{B}V_g\widehat{B}^\top -BV_g B^\top \|_{\max}
\|\widehat{w}_{mv}\|_{\max}
\\ &
+\|{w}_{mv}^*\|_{\max}\|BV_gB^\top\|_{\max}  \|\widehat{w}_{mv} - w_{mv}^* \|_{\max}.
\end{align*}

By Assumption 7(a), $\| B\|_{\max}$ is bounded, and hence $\| \widehat{B}\|_{\max}$ is also bounded in probability by the relationship $\|\widehat{B}\|_{\max} \leq \|{B}\|_{\max} + \|\widehat{B} - B\|_{\max} $ where $\|\widehat{B} - B\|_{\max} \leq \max_i \| \widehat{b}_i - b_i\| = O_p(\omega_T) = o_p(1)$ by~\eqref{Lemma: asy for factor model}. Next, for mean-variance problem (1.1), its solution $w^{*}_{mv}$ has the closed form as follow:
\begin{equation}
w_{mv}^{*} = \frac{\bar{\alpha} A_3 -A_1}{A_4}\Sigma_{r}^{-1}\mu + \frac{A_2 - \bar{\alpha }A_1}{A_4}\Sigma_{r}^{-1}\bm{1}_p,\label{eq: optimal weight}
\end{equation}
where $\bar{\alpha}$ is a fixed constant denoting the expected portfolio return,
$A_1 = \mu^\top\Sigma_{r}^{-1}\bm{1}_p$, $A_2 = \mu^\top \Sigma_{r}^{-1}\mu$,  $A_3 = \bm{1}_p^\top \Sigma_{r}^{-1}\bm{1}_p$ and $A_4 = A_2A_3- A_1^2$.  Then, it follows that 
\begin{equation}
\|w_{mv}^*\|_{\max} \leq |\bar{\alpha} A_3-A_1||A_4^{-1}|\| \Sigma_r^{-1}\mu\| + |A_2 - \bar{\alpha} A_1 ||A_4^{-1}|\| \| \Sigma_r^{-1}\bm{1}_p\|
= O_p\left( {p^{1/2-\phi}}\right).\label{eq: w star rate}
\end{equation}

Now we derive for $\| \widehat{w}_{mv}-w_{mv}^*\|_{\max}$, it follows that 
\begin{align*}
\widehat{w}_{mv}- w_{mv}^{*}=& 
\frac{\bar{\alpha} \widehat{A}_3-\widehat{A}_1}{\widehat{A}_4}\widehat{\Sigma}_{r}^{-1}\widehat{\mu} +\frac{\widehat{A}_2- \bar{\alpha}\widehat{A}_1}{\widehat{A}_4}\widehat{\Sigma}_{r}^{-1}\bm{1}_p
-\frac{\bar{\alpha} A_3-A_1}{A_4}\Sigma_{r}^{-1}\mu - \frac{A_2- \bar{\alpha}A_1}{A_4}\Sigma_{r}^{-1}\bm{1}_p \\ 
=&\left(\frac{\bar{\alpha} \widehat{A}_3-\widehat{A}_1}{\widehat{A}_4} - \frac{\bar{\alpha} A_3-A_1}{A_4}\right)\widehat{\Sigma}_{r}^{-1}\widehat{\mu} + \frac{\bar{\alpha} A_3-A_1}{A_4}\left( \widehat{\Sigma}_{r}^{-1}\widehat{\mu} - \Sigma_r^{-1}\mu\right)\\ 
& + \left(\frac{\widehat{A}_2- \bar{\alpha}\widehat{A}_1}{\widehat{A}_4} - \frac{A_2- \bar{\alpha}A_1}{A_4}\right)\widehat{\Sigma}_{r}^{-1}\bm{1}_p + \frac{A_2- \bar{\alpha}A_1}{A_4} \left( 
 \widehat{\Sigma}_{r}^{-1}\bm{1}_p - \Sigma_{r}^{-1}\bm{1}_p\right)\\ 
=& \mathcal{I}_1 + \mathcal{I}_2 + \mathcal{I}_3 + \mathcal{I}_4.
\end{align*}
For $\mathcal{I}_{1}$, we have 
\begin{align*}
\| \mathcal{I}_1\| &\leq \left| \frac{\bar{\alpha} \widehat{A}_3-\widehat{A}_1}{\widehat{A}_4} - \frac{\bar{\alpha} A_3-A_1}{A_4}\right| \|\widehat{\Sigma}_{r}^{-1}\widehat{\mu}\|\\
& = \left|\frac{\bar{\alpha}\left(\widehat{A}_3A_4 - A_3\widehat{A}_4\right) + (A_1\widehat{A}_4 - \widehat{A}_1A_4)}{A_4\widehat{A}_4}\right| \|\widehat{\Sigma}_{r}^{-1}\widehat{\mu}\|\\ 
&=\left|\frac{\bar{\alpha}\left(\widehat{A}_3A_4 - A_3A_4 + A_3A_4- A_3\widehat{A}_4\right) + (A_1\widehat{A}_4 - A_1A_4 + A_1A_4 - \widehat{A}_1A_4)}{A_4\widehat{A}_4}\right| \|\widehat{\Sigma}_{r}^{-1}\widehat{\mu}\|\\
& \leq \left(\left|\frac{\bar{\alpha}(\widehat{A}_3 - A_3)}{\widehat{A}_4}\right| + \left|\frac{\bar{\alpha}A_3(A_4-\widehat{A}_4)}{A_4\widehat{A}_4} \right| + \left| \frac{A_1(\widehat{A}_4-A_4)}{A_4\widehat{A}_4}\right| +\left| \frac{A_1-\widehat{A}_1}{\widehat{A}_4}\right|\right)\|\widehat{\Sigma}_{r}^{-1}\widehat{\mu} \|.
\end{align*}
Note that 
\begin{align}
\left|A_1 - \widehat{A}_1 \right| & =  \left| \mu^\top\Sigma_{r}^{-1}\bm{1}_p -\widehat{\mu}^\top\widehat{\Sigma}_{r}^{-1}\bm{1}_p \right| \notag\\ 
& = \left| \mu^\top\Sigma_{r}^{-1}\bm{1}_p - \widehat{\mu}^\top\Sigma_{r}^{-1}\bm{1}_p + \widehat{\mu}^\top\Sigma_{r}^{-1}\bm{1}_p - \widehat{\mu}^\top\widehat{\Sigma}_{r}^{-1}\bm{1}_p \right| \notag\\ 
& \leq \| \widehat{\mu}-\mu\|_{\max} \| \Sigma_r^{-1}\bm{1}_p \| + \| \widehat{\Sigma}_r^{-1} -\Sigma_r^{-1}\| \|\bm{1}_p \|_{\max} \|\widehat{\mu}\|_{\max} \notag\\ 
& = O_p\left( p^{1/2}\zeta(p,T)+ 
m_\varsigma \omega_T^{1-\varsigma}\right), \label{eq: A1 -hat A1}
\end{align}
where $\zeta(p,T) = \sqrt{\frac{\log p}{T}} + \frac{T^{1/4}}{\sqrt{p}}$,  $\|\widehat{\Sigma}_{r}^{-1}-\Sigma_r^{-1} \|= O_p(m_\varsigma \omega_T^{1-\varsigma})$ by Lemma~\ref{lemma fan2013 theorem 3.1 .32}, and 
\begin{align}
\| \widehat{\mu} - \mu\|_{\max} & = \| \widehat{B}\widehat{\mu}_f - B\mu_f\|_{\max} \\& \leq \| \widehat{B} - B\|_{\max} \|\mu_f\|_{\max} + \|\widehat{\mu}_f -\mu_f\| \|B\|_{\max} \notag\\ &
= O_p\left( \omega_T + \left(\frac{1}{\sqrt{T}}+\frac{T^{1/4}}{\sqrt{p}}\right)\right) \notag \\& = O_p\left(\sqrt{\frac{\log p}{T}} +  \frac{T^{1/4}}{\sqrt{p}}\right), \label{eq: hatmu -mu}
\end{align}
where the order of $\|\widehat{\mu}_f - \mu_f\|$ is given by \eqref{eq: appendix rate hatmu -mu}, $\| \widehat{\mu}\|_{\max}$ is bounded in probability given the condition {$T= o(p^2)$} and $\|\mu\|_{\max} < M$ for some positive constant $M$ indicated by Assumptions 1(b) and 7(a). Similarly, we can also derive that
\begin{align}
|A_2 - \widehat{A}_2| &= \left| \mu^\top\Sigma_{r}^{-1}\mu -\widehat{\mu}^\top\widehat{\Sigma}_{r}^{-1}\widehat{\mu} \right| \notag\\ & 
    = \left|\mu^\top\Sigma_{r}^{-1}\mu - \mu^\top\widehat{\Sigma}_{r}^{-1}\mu + \mu^\top\widehat{\Sigma}_{r}^{-1}\mu - \widehat{\mu}^\top\widehat{\Sigma}_{r}^{-1}{\mu} + \widehat{\mu}^\top\widehat{\Sigma}_{r}^{-1}{\mu} -  \widehat{\mu}^\top\widehat{\Sigma}_{r}^{-1}\widehat{\mu} \right| \notag
    \\ & = 
    \left| \mu^\top(\Sigma_{r}^{-1} - \widehat{\Sigma}_r^{-1})\mu  +  (\mu^\top -\widehat{\mu}^\top)\widehat{\Sigma}_{r}^{-1}\mu + \widehat{\mu}^\top\widehat{\Sigma}_{r}^{-1}(\mu-\widehat{\mu})
    \right| 
    \notag
    \\ &
    = O_p\left(  m_\varsigma \omega_T^{1-\varsigma} +  p^{1/2}\zeta(p,T) \right),
    \label{eq:A2-hatA2}
\end{align}

Next, 
\begin{equation}
|A_3 - \widehat{A}_3| = \left| \bm{1}_p^\top\Sigma_{r}^{-1}\bm{1}_p -\bm{1}_p^\top\widehat{\Sigma}_{r}^{-1}\bm{1}_p \right| \leq \|\bm{1}_p\|_{max}\| \Sigma_r^{-1} - \widehat{\Sigma}_r^{-1}\|\|\bm{1}_p\|_{\max} = O_p\left(m_\varsigma \omega_T^{1-\varsigma}\right).\label{eq: A3- hat A3}
\end{equation}
Under the condition that {$p^{1/2-\phi}\zeta(p,T) + p^{-\phi}m_\varsigma \omega_T^{1-\varsigma} = o(1)$} indicated by Assumption 10(c), it follows that $\widehat{A}_i = O_p(p^\phi)$ for $i = 1,2,3$, thus we have 
\begin{align}
|A_4 -\widehat{A}_4| &= |A_2A_3 - \widehat{A}_2\widehat{A}_3 +  \widehat{A}_1^2 - {A}_1^2| \notag\\ &
=|A_2A_3 - \widehat{A}_2A_3 + \widehat{A}_2A_3 -\widehat{A}_2\widehat{A}_3 +  (\widehat{A}_1 - {A}_1)(A_1 + \widehat{A}_1)| \notag\\ &
\leq |A_2-\widehat{A}_2||A_3| + |A_3 -\widehat{A}_3||\widehat{A}_2| + |\widehat{A}_1 - {A}_1||A_1 + \widehat{A}_1| \notag\\ & = O_p\left( p^\phi m_\varsigma \omega_T^{1-\varsigma} + p^{1/2+\phi}\zeta(p,T)\right)
\label{eq: A4 - hat A4}
\end{align}
where we apply Eq. \eqref{eq: A1 -hat A1}, \eqref{eq:A2-hatA2}, \eqref{eq: A3- hat A3},  and Assumptions 10. Note that $\widehat{A}_4 = A_4 + (\widehat{A}_4 - A_4)$, the conditions {$p^{-\phi}m_\varsigma\omega_T^{1-\varsigma} +  p^{1/2-\phi}\zeta(p,T) = o(1)$} (indicated by Assumption 10(c)) ensures  that $\widehat{A}_4 \asymp p^{2\phi}$. 
 As a result, we further obtain 
\begin{align}
\| \mathcal{I}_1\| & \leq 
    \left(\left|\frac{\bar{\alpha}(\widehat{A}_3 - A_3)}{\widehat{A}_4}\right| + \left|\frac{\bar{\alpha}A_3(A_4-\widehat{A}_4)}{A_4\widehat{A}_4} \right| + \left| \frac{A_1(\widehat{A}_4-A_4)}{A_4\widehat{A}_4}\right| +\left| \frac{A_1-\widehat{A}_1}{\widehat{A}_4}\right|\right)\|\widehat{\Sigma}_{r}^{-1}\widehat{\mu} \| \notag
\\ & 
    = O_p\left( 
    p^{1-2\phi}\zeta(p,T) + p^{1/2-2\phi}m_\varsigma \omega_T^{1-\varsigma}
    \right)
\label{eq: I1 rate}
\end{align}
Then, consider $\mathcal{I}_2$,
\begin{align}
 \| \mathcal{I}_2\|_{max} &= \left| \frac{\bar{\alpha} A_3-A_1}{A_4}\right|   \|  \widehat{\Sigma}_{r}^{-1}\widehat{\mu} - \Sigma_r^{-1}\mu \|_{max} \notag
 \\ & \leq 
 \left| \frac{\bar{\alpha} A_3-A_1}{A_4}\right| \left(   \|  \widehat{\Sigma}_{r}^{-1}\widehat{\mu} - \Sigma_r^{-1}\widehat{\mu} \|_{max} + \|\Sigma_r^{-1}\widehat{\mu} - \Sigma_r^{-1}{\mu} \|_{max} \right) \notag
 \\&
 \leq 
 \left| \frac{\bar{\alpha} A_3-A_1}{A_4}\right| \left(   \|  \widehat{\Sigma}_{r}^{-1} - \Sigma_r^{-1}\| \|\widehat{\mu} \|_{max} +  \| \Sigma_r^{-1}\| \|\widehat{\mu} - {\mu} \|_{max} \right) \notag
 \\ & 
 = O_p  \left(p^{-\phi}m_\varsigma \omega_T^{1-\varsigma} + p^{-\phi}\zeta(p,T)\right) 
 \label{eq: I2 rate}
\end{align}
where $\|\Sigma_r^{-1}\|$ is bounded since $\lambda_{\min}(BB^\top) = 0$, $c_2 > \lambda_{\min}(\Sigma_e) > c_1$ and Weyl's Theorem. Next, 
\begin{align}
\| \mathcal{I}_3\| &\leq \left(\left|\frac{\widehat{A}_2-A_2}{A_4}\right| + \left|\frac{A_2(A_4-\widehat{A}_4)}{\widehat{A}_4 A_4} \right| + \left| \frac{\bar{\alpha}A_1(\widehat{A}_4 - A_4)}{\widehat{A}_4 A_4}\right| + \left| \frac{\bar{\alpha}(A_1-\widehat{A}_1)}{\widehat{A}_4} \right|
\right)\|\widehat{\Sigma}_{r}^{-1}\bm{1}_p \| \notag
\\ &
=O_p\left( 
    p^{1-2\phi}\zeta(p,T) + p^{1/2-2\phi}m_\varsigma \omega_T^{1-\varsigma}
    \right).
\label{eq: I3 rate}
\end{align}
Finally, we have 
\begin{align}
\| \mathcal{I}_4 \|_{\max} = \left| \frac{A_2- \bar{\alpha}A_1}{A_4} \right| \left\|
\widehat{\Sigma}_{r}^{-1} - \Sigma_{r}^{-1}\right\| \| \bm{1}_p\|_{\max} = O_p\left(p^{-\phi}m_\varsigma \omega_T^{1-\varsigma} \right).
\label{eq: I4 rate}
\end{align}
Combine~\eqref{eq: I1 rate},~\eqref{eq: I2 rate},~\eqref{eq: I3 rate}, and~\eqref{eq: I4 rate}, we have
\begin{equation}
\|\widehat{w}_{mv} - w_{mv}^*\|_{\max} = O_p\left( (p^{1-2\phi} + p^{-\phi})\zeta(p,T)+ (p^{1/2-2\phi}+p^{-\phi})m_\varsigma \omega_T^{1-\varsigma}\right).
\label{eq: hat w - w rate}
\end{equation}

Based on~\eqref{eq: w star rate},~\eqref{eq: hat w - w rate}, and the fact that $\|B\|_{max}$ and $\|V_g\|$ are bounded, it follows that 
\begin{align}
\|{w}_{mv}^*\|_{\max}\|BV_gB^\top\|_{\max}  \|\widehat{w}_{mv} - w_{mv}^* \|_{\max} &= O_p\left( (p^{3/2-3\phi}+p^{1/2-2\phi})\zeta(p,T) + (p^{1-3\phi}+p^{1/2-2\phi})m_\varsigma \omega_T^{1-\varsigma} \right) \notag\\ 
&= o_p(1)
\end{align}
given $p^{3/2-3\phi}\zeta(p,T) + p^{1-3\phi}m_\varsigma \omega_T^{1-\varsigma} = o(1)$. By similar arguments, we also have that $
\|\widehat{w}_{mv} - w_{mv}^{*}\|_{\max} \| \widehat{B}V_g\widehat{B}^\top\|_{\max} \|\widehat{w}_{mv}\|_{\max} = o_p(1)$.
Finally,
\begin{align}
&\|{w}_{mv}^*\|_{\max}\| \widehat{B}V_g\widehat{B}^\top -BV_g B^\top \|_{\max}
\|\widehat{w}_{mv}\|_{\max} \notag \\&= \|{w}_{mv}^*\|_{\max}\| \widehat{B}V_g\widehat{B}^\top - {B}V_g\widehat{B}^\top+ {B}V_g\widehat{B}^\top -  BV_g B^\top \|_{\max}
\|\widehat{w}_{mv}\|_{\max} \notag
\\ &
\leq \|{w}_{mv}^{*}\|_{\max}\left(\| \widehat{B} - B\|_{\max} \|V_g\| \|\widehat{B}\|_{\max} + \|V_g\| \|{B}\|_{\max} \|\widehat{B}^\top -  B^\top \|_{\max} \right)
\|\widehat{w}_{mv}\|_{\max} \notag
\\ &
= O_p\left( p^{1-2\phi}\omega_T \right) = o_p(1),
\end{align}
where we apply~\eqref{eq: w star rate},~\eqref{eq: hat w - w rate}, and the fact that $\|\widehat{B}-B\|_{\max} = O_{p}(\omega_T)$ for the last inequality. $p^{1-2\phi} \omega_T = o(1)$ is implied by $p^{3/2-3\phi} \zeta(p,T) = o(1)$  and $\omega_T/\zeta(p,T) = O(1)$. Consequently, we can draw the conclusion that $ {w}_{mv}^{*\top} {B} {V}_{g} {B}^{\top} {w}_{mv}^{*} = \widehat{w}_{mv}^{\top}\widehat{B}{V}_{g} \widehat{B}^{\top}\widehat{w}_{mv} + o_p(1)$.

The remaining task is to show that replacing $V_g$ with $\widehat{V}_{g,T}$ does not affect the asymptotic results, that is 
\begin{align}
\left|\widehat{w}_{mv}^\top \widehat{B} {V}_{g} \widehat{B}^\top \widehat{w}_{mv} - \widehat{w}_{mv}^\top \widehat{B} \widehat{V}_{g,T} \widehat{B}^\top \widehat{w}_{mv} \right|&\leq C \| \widehat{w}_{mv}\|_{\max}^2 \|\widehat{B}\|_{\max}^2 \|V_g -\widehat{V}_{g,T} \| \notag
\\&  = O_p\left( p^{1-2\phi}q_T\left(\frac{1}{\sqrt{T}} + \frac{T^{1/4}}{\sqrt{p}}\right) \right) = o_p(1)
\end{align}
where the order of $\|\widehat{V}_g -V_g\| = O_p\left(q_TT^{-1/2} + q_TT^{1/4}p^{-1/2}\right)$ is from the proof of Theorem 4, and the last equality holds by Assumption 10(c).

\end{document}